\documentclass[aps,%
 reprint,
 amsmath,amssymb,
  nofootinbib
]{revtex4-2}

\usepackage{graphicx}
\usepackage{dcolumn}
\usepackage{bm}

\usepackage{booktabs}
\usepackage{tikz}
\usepackage{xcolor}

\usepackage{lpic}
\usepackage{subcaption}
\usepackage{dsfont}

\usepackage{booktabs}
\usepackage{multirow}
\usepackage{mathtools}
\usepackage{hyperref}
\usepackage[shortlabels]{enumitem}
\usepackage{braket}
\definecolor{dlrdarkgrey}{RGB}{121, 122, 121}
\definecolor{dlrmedgrey}{RGB}{154, 155, 154}
\definecolor{dlrlightgrey}{RGB}{230, 231, 230}
\definecolor{dlrblue}{RGB}{12, 102, 137}
\definecolor{dlrgreen}{RGB}{149, 178, 62}
\definecolor{dlryellow}{RGB}{238, 201, 47}

\begin{document}

\preprint{APS/123-QED}

\title{Effects of pulse durations in atom interferometry with double Raman diffraction}

\author{Nadja Augst}
\author{Albert Roura}
\affiliation{%
German Aerospace Center (DLR), Institute of Quantum Technologies,\\
Wilhelm-Runge-Straße 10, 89081 Ulm, Germany
}%

\date{\today}

\begin{abstract}
Double Raman diffraction is a key enabling technique for future space missions relying on atom interferometry that offers enhanced sensitivity and intrinsic suppression of systematic effects. This diffraction mechanism naturally arises when employing a retro-reflection configuration in microgravity, but then accelerations can no longer be determined through a frequency-chirp scan. As a consequence, measurements become much more sensitive to systematic effects associated with pulse timing and duration. Indeed, the finite pulse duration can introduce additional phase shifts and small modifications of the arm trajectories, which are critical sources of error in high-precision measurements. In this work, we present the first complete analytical results of these effects for double Raman diffraction, building upon previous work for single diffraction and extending the standard literature. Our analysis determines the impact of finite pulse durations on the full interferometer sequence, quantifying the shifts of the atomic wave packets' trajectories and the acquired phases. The description relies on a comprehensive framework describing matter-wave propagation in terms of central trajectories and centered wave packets, and can be easily incorporated into relativistic treatments. Furthermore, we make use of group theoretic methods that are particularly useful when accounting for residual accelerations and frequency chirps, which are also relevant for ground-based applications. Our study includes time-dependent accelerations and frequency chirps, gravity gradients, retardation effects, AC Stark shifts induced by the Raman pulses, and external forces such as those due to magnetic fields or temperature gradients in the black-body radiation.
\end{abstract}

\maketitle

\section{Introduction}
\subsection{Context and motivation}
\label{sec:intro}
\begin{figure}[t]
    \centering
    \includegraphics[width=.53\textwidth]{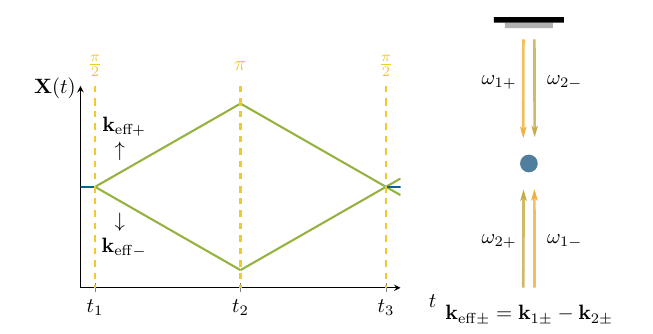}
    \caption{Schematic diagram of a light-pulse atom interferometer based on double Raman diffraction. (Left) Symmetric but opposite momentum transfers are imparted to the two arms of the interferometer. In the co-moving frame, this leads to identical phase evolution along both paths, resulting in symmetric phase terms that cancel out in the final interferometric phase. (Right) Retro-reflection enables cancellation of laser phase noise for short baselines. Each frequency component $\omega$ corresponds to a wave vector $\mathbf{k}$, yielding an effective momentum transfer $\mathbf{k}_\mathrm{eff}$; the $+$ and $-$ signs denote quantities associated with the two opposite pairs of counter-propagating laser beams. For terrestrial applications, three frequency components are required to compensate the Doppler shift.}
    \label{fig:double_raman_intro}
\end{figure}

Light-pulse atom interferometers \cite{Kasevich1991,Kleinert_2015} are excellent quantum inertial sensors \cite{Peters_1999,Savoie_2018} for both practical applications and fundamental physics measurements \cite{Bongs2019}. These include the accurate determination of fundamental constants \cite{Rosi2014,Parker_2018,Morel2020} as well as high-precision tests of QED \cite{Bouchendira_2011,Parker_2018,Morel2020}, the equivalence principle \cite{Schlippert2014,Zhou_2015,Rosi2017,Asenbaum2020}, and certain dark-energy models \cite{Hamilton2015,Jaffe_2017,Sabulsky_2019}.
A widely used mechanism for coherent momentum transfer in light-pulse atom interferometry are two-photon processes such as Bragg and Raman diffraction. In Bragg diffraction the internal state remains unchanged, while in Raman diffraction the atoms undergo a transition between different hyperfine ground states, enabling state-selective detection. Both schemes are well-suited for high-precision measurements in space and terrestrial environments alike.\\
\\
Double diffraction involves two reflection-symmetric pairs of counter-propagating laser beams that generate opposite momentum transfers in both arms of the interferometer, as illustrated in Fig.\,\ref{fig:double_raman_intro}. The use of a retro-reflection mirror ensures cancellation of laser phase noise for short baselines (although this cancellation does not extend to mechanical vibrations of the mirror, which remains a noise source). In microgravity environments, atoms are naturally released at rest, resulting in simultaneous resonant coupling to both pairs of counter-propagating laser beams
\footnote{For atoms in magnetically sensitive states this can be avoided through a suitable choice of laser polarizations \cite{Bernard_2022}. Alternatively, it can also be avoided by injecting two laser frequency components with an off-resonant frequency difference combined with a larger frequency chirp during the pulse that exploits the finite time-of-flight to the retro-reflection mirror \cite{Perrin_2019b}. But these methods target compact set-ups rather than highly accurate measurements in microgravity.}.
Atom interferometers based on double diffraction have been experimentally demonstrated for both Raman \cite{Leveque_2009} and Bragg diffraction \cite{Ahlers_2016,Kubler2016}, with double Raman diffraction offering additional advantages in terms of state control.\\
\\
Compared to Bragg, off-resonant transitions are largely suppressed in the Raman case (with detunings of several GHz corresponding to the hyperfine splitting) and the associated parasitic paths, which can be an important source of systematic uncertainties \cite{Jenewein2022,Siemss_2023}, are avoided. Furthermore, atoms left undiffracted by the first beam splitter or the mirror pulses can be removed with blow-away pulses \cite{Leveque_2009}, thereby improving interferometric contrast and measurement accuracy. This is the main reason why space missions such as BECCAL \cite{Frye2021}, CARIOQA \cite{Leveque2023}, MAQuIS \cite{Woerner2023} and STE-QUEST \cite{Aguilera2014, Struckmann2024} plan to employ double Raman diffraction, a key motivation for studying it in this work.\\
\\
Double diffraction is also relevant for terrestrial applications. In horizontal configurations, the absence of gravitational acceleration along the direction of the laser beams mimics microgravity conditions. In vertical configurations, on the other hand, a frequency chirp is required to compensate for gravitational acceleration, which is only possible in retro-reflection set-ups by adding a third frequency component \cite{Malossi_2010,Zhou_2015}. In our analysis, we include the possibility of a frequency chirp in the laser frame to account for such situations. A motivation for using double diffraction in atom interferometers based on Raman transitions (even ground-based ones) is the insensitivity to time-dependent homogeneous magnetic fields, which induce different energy shits for the two hyperfine states \cite{He2021}. Indeed, contrary to single Raman diffraction, in double diffraction both arms experience the same internal-state evolution, and such effects cancel out. \\
\\
Terrestrial atom interferometers typically mitigate systematic effects from finite pulse durations by scanning the laser chirp shot-to-shot. However, for double diffraction this approach is incompatible with retro-reflective geometries involving just two frequency components. Therefore, a precise analytical understanding of pulse-duration-induced phase shifts becomes essential for accurate measurements. While such effects have been characterized for single diffraction \cite{Antoine_2006,Bertoldi2019, Bonnin_2015, Saywell2026}, their impact in double-diffraction schemes remains unaddressed.
\subsection{Approach employed in this work}
\label{sec:intromethods}
  We make use of the centered wave-packet formalism, decomposing the state of the atom's center-of-mass degree of freedom into a central trajectory, which follows the classical equations of motion, and a centered wave packet, which describes its expansion and shape evolution. For free evolution, the wave packet can be expressed \cite{Augst2027} as follows:
  \begin{equation}
  \label{eq:centereedWPfree}
      \psi(\mathbf{x}, t) =  e^{\frac{i}{\hbar} \mathcal{S}} e^{ \frac{i}{\hbar}\mathbf{P}  \cdot \left( \mathbf{x} - \mathbf{X}(t) \right)}  \psi_\mathrm{c}(\mathbf{x}-\mathbf{X}(t),t) \,,
  \end{equation}
  where \(\mathbf{X}(t)\) is the central trajectory $\mathbf{P}$ and is the central m,omentum, $\psi_\mathrm{c}(\mathbf{x}',t)$ is centered wave packet , and the spatially independent propagation phase \(\mathcal{S}\) corresponds to the classical action evaluated along \(\mathbf{X}(t)\). In the context of the centered wave-packet formalism, capitalized variables denote quantities associated with the central trajecotry. The total momentum is decomposed into the central component $\mathbf{P}$ and the relative momentum $\mathbf{p}'$ as
  \begin{equation}
  \label{eq:momentumdecomposition}
      \mathbf{p} = \mathbf{P} + \mathbf{p}' \,.
  \end{equation}
  The centered wave-packet formalism has been generalized to relativistic settings including curved spacetime \cite{Dimopoulos2008,Roura2020,Swan_2026}.\\
  \\
  When considering residual accelerations in microgravity, accelerated setups, or uniform gravitational fields, it is convenient to work in the freely falling frame co-moving with the mid-point trajectory: in this reference frame, central trajectories are straight lines, ensuring symmetry between interferometer arms. Many phase terms are identical in both arms and cancel out in the phase difference. Symmetry is maintained even in the presence of uniform gravitational gradients, despite the curvature of the central trajectories.\\
  \\
  When analyzing the time evolution during the pulses, it is convenient to work in the momentum representation, where the atomic wave packets involve a superposition of different momentum eigenstates. The reason is that the evolution becomes much simpler for momentum eigenstates (analogously to the case of free evolution). In particular, the dynamics can be organized in terms of small Hilbert subspaces that are associated with a set of two (single diffraction) or three (double diffraction) momentum eigenstates, and each subspace evolves independently.\\
  \\
  We analyze two- and three-level systems using group-theoretic methods: Single diffraction, involving the two-level system \( \ket{\mathrm{g}, \mathbf{p}}\) and \(\ket{\mathrm{e}, \mathbf{p} + \hbar \mathbf{k}_\mathrm{eff}}\) in ground and excited states  respectively, is described by the spin-1/2 representation, corresponding to the fundamental representation of SU(2). For two-photon processes, the intermediate excited state is adiabatically eliminated \cite{Steck2020, Bernhardt_1981, Marte_1992, Brion_2007}, yielding an effective two-level system characterized by the two-photon frequency difference
    \begin{equation}
        \Delta \omega_{\pm} = \omega_{1\pm} - \omega_{2\pm} \,,
    \end{equation}
    an effective laser phase
    \begin{equation}
        \varphi_\pm = \varphi_{1\pm} - \varphi_{2\pm} \,,
    \end{equation}
    and the effective wave vector
    \begin{equation}
        \mathbf{k}_{\mathrm{eff}\pm} = \mathbf{k}_{1\pm} - \mathbf{k}_{2\pm} \,,
    \end{equation}
    as illustrated in Fig.~\ref{fig:double_raman_intro}. The $\pm$ labels denote quantities associated with each one of the two opposite pairs of counter-propagating laser beams. In a retroreflective geometry, symmetry enforces $\Delta \omega_+ = \Delta \omega_-$ and $\mathbf{k}_{\mathrm{eff}+} = -\mathbf{k}_{\mathrm{eff}-} \equiv \mathbf{k}_{\mathrm{eff}}$.\\
    \\
  Double diffraction, described by the three-level system \(\ket{\mathrm{g}, \mathbf{p}}\), \(\ket{\mathrm{e}, \mathbf{p} + \hbar \mathbf{k}_\mathrm{eff}}\) and \(\ket{\mathrm{e}, \mathbf{p} -\hbar \mathbf{k}_\mathrm{eff}}\), from now on abbreviated as \(\ket{\mathrm{g}}\) and \(\ket{\pm}\), respectively, where \(\pm\) denotes the positive- and negative-recoil branches. This three-level system will correspond to the spin-1 representation of the SU(2) group. \\
  \\
  The generators of the SU(2) group can be written as $(\hat{J}_l/\hbar)$ in terms of the angular momentum operators $\hat{J}_l$ and are characterized by the commutation relation $[\hat{J}_i, \hat{J}_j] = i\hbar \, \varepsilon_{ijk} \hat{J}_k$. The time evolution is governed by the operator $\exp\left(-i \Omega_\mathrm{eff} t \, n^l \hat{J}_l/\hbar \right)$, where $n^l$ is the direction of the effective coupling, and the effective Rabi frequency is given by 
  \begin{equation}
  \label{eq:omegaeff}
      \Omega_\mathrm{eff} = \sqrt{\Omega_\mathrm{D}^2 + \delta^2} \,,
  \end{equation}
    with the generalized Rabi frequency $\Omega_\mathrm{D}$ for a three-level system, and the detuning $\delta$ defined below.\\
  \\
  The generators $(\hat{J}_i / \hbar)$ have different matrix representations for spin~1/2 and spin~1. For the spin-1/2 representation they are typically expressed in terms of the Pauli matrices as $J_i/\hbar =  \sigma_i/2$. In this case, the identity $(n^l \sigma_l)^2 = 1$ leads to the following generalized Euler formula (see Ref.\,\cite{Augst2027}):
  \begin{equation}
      e^{-i \frac{\Omega_\mathrm{eff} t}{2} n^l \sigma_{l} } = \cos\left(\frac{\Omega_\mathrm{eff} t}{2}\right) \mathbf{I} - (i n^l \sigma_{l})\, \sin \left(\frac{ \Omega_\mathrm{eff} t}{2}\right) \,.
  \end{equation}
  In contrast, the simple identity for the square of the generators does not hold for the spin-1 representation, but in that case the relation $(n^l \hat{J}_l / \hbar)^3 = n^l \hat{J}_l/\hbar$ allows the derivation of Rodrigues’ formula (see App.\,\ref{app:SU2}):
  \begin{align}
  \label{eq:eulerspin1}
      e^{-i \Omega_\mathrm{eff} t \, n^l \hat{J}_l/\hbar }  &= \mathbf{I}- i n^l \hat{J}_l/\hbar \sin \left(\Omega_\mathrm{eff} t \right) \notag \\
      &+ (n^l \hat{J}_l/\hbar)^2 (\cos \left(\Omega_\mathrm{eff} t \right)-1)\,,
  \end{align}
  As seen above, for any momentum eigenstate the evolution depends on the detuning, defined as
  \begin{equation}
  \label{eq:detuning}
      \delta_\pm = (E_\pm - E_0) / \hbar - \Delta\omega_\pm \,,
  \end{equation}
  where the energies $E_0 = E_\mathrm{g} +  \mathbf{p}^2/(2m)$ and $E_{\pm} = E_\mathrm{g}  + \hbar \omega_\mathrm{eg} +  (\mathbf{p} \pm \hbar \mathbf{k}_\mathrm{eff})^2/(2m) $ include internal and kinetic energy. The detuning can be decomposed into a $\mathbf{p}'$-independent term $\delta_\pm^{(\mathrm{c})}$, which depends on the central momentum $\mathbf{P}$, and a $\mathbf{p}'$-dependent part $\delta_\pm'(\mathbf{p}')$.\\
  \\
  For wave packets with a sufficiently narrow momentum width, the evolution involves a time-dependent amplitude and a phase factor whose phase consists of terms independent of $\mathbf{p}'$ as well as $\mathbf{p}'$-dependent ones. The latter include terms linear in $\mathbf{p}'$ and quadratic in $\mathbf{p}'$. The linear terms lead to a shift of the central trajectory (through the Fourier shift theorem), and the quadratic ones imply an evolution of the centered wave packet that is actually the same as if it were freely evolving. The terms independent of $\mathbf{p}'$, on the other hand, can be systematically organized in such a way that they coincide with the phase that would be accumulated by a wave packet that were freely evolving in the initial state up to a time $t_j$, when a laser pulse with vanishing duration is applied, and after that freely evolved again in the final state. This intuitive way of organizing the phase contributions as free propagation phases and laser phases at time $t_j$ can be diagrammatically represented with corresponding \emph{mnemonic trajectories}.
  In addition, there are phase corrections associated with the non-vanishing pulse duration that mainly depend on the detunings $\delta_\pm^{(\mathrm{c})}$ for the central momentum $\mathbf{P}$.\\
  \\
  Note that although there is freedom concerning the actual values of $t_j$ that one takes for each pulse, observable quantities such as the total phase shift for the full interferometer are independent of the actual choice. For example, a convenient choice that will be considered in the next sections are the times $t_j^*$ for which the mnemonic trajectories coincide at the beginning and the end of the pulse with the actual central trajectories.\\
  \\
   In the freely falling frame, it is particularly convenient to consider the symmetric and antisymmetric linear combinations. For the detuning this corresponds to the following symmetric 
  \begin{equation}
  \label{eq:detuningsymm}
      \delta_\mathrm{S} = (\delta_+ + \delta_-)/2
  \end{equation}
  and antisymmetric 
  \begin{equation}
  \label{eq:detuningantisymm}
      \delta_\mathrm{A} = (\delta_+ - \delta_-)/2
  \end{equation}
  parts. Doppler shifts affect only the antisymmetric part. On the other hand, under appropriate conditions discussed in Sec.~\ref{sec:AC-Stark-shift}, the symmetric part $\delta_\mathrm{S}$ is identical in both arms, and the corresponding phase-shift contributions cancel out. There is a completely analogous decompositon into symmetric and antisymmetric parts for the two-photon frequency differences $\Delta \omega_\pm$, the energies $E_\pm$ and the frequency-chirp rates $\alpha_\pm$.

\subsection{Outline and notation}

    The remainder of this paper is organized as follows: In Sec.~\ref{sec:single_pulse}, we investigate the effects of finite pulse duration in single pulses, introducing a systematic organization of phase terms. Sec.~\ref{sec:fullAI} extends this analysis to the full interferometer, examining both closed and open configurations and the transition between the laboratory and freely falling frames. This transition naturally motivates the inclusion of residual accelerations and uncompensated frequency chirps, which are treated in Sec.~\ref{sec:chirpgeneral}. Finally, Sec.~\ref{sec:additional_effects} generalizes the formalism to time-dependent accelerations and chirps, gravity gradients, asymmetric chirps, retardation effects, and various experimental imperfections, including magnetic fields and other external forces, AC Stark shifts, and Gaussian beam profiles.\\ 
    \\
    Throughout this work, vector quantities are denoted in boldface (e.g., $\mathbf{p}$, $\mathbf{k}$), while scalar quantities are written in regular (non-bold) type. Indices written in italic (cursive) are interchangeable and may be summed via the Einstein summation convention when repeated (e.g., $n^l J_l$). In contrast, roman (upright) indices (e.g., $t_\mathrm{i}$, $\mathbf{k}_\mathrm{eff}$) are used to denote distinct physical entities and carry fixed meaning. Operators are denoted with a hat ($\hat{H}$, $\hat{U}$), unless expressed in basis-dependent matrix form.

\section{Single pulses}
\label{sec:single_pulse}
Trajectory and phase during a square pulse are described within the centered wave-packet formalism (see \cite{Augst2027}). The evolution during the pulse is given by
\begin{align}
\label{eq:psigeneral}
    \psi^{(r)}(\mathbf{x}, t) &=  e^{i \phi^{(j)}_{rs}} \,  e^{i \phi^{(\mathrm{sq})}_{rs}} \, e^{\frac{i}{\hbar} \mathcal{S}_{rs}} \,e^{ \frac{i}{\hbar}\mathbf{P}_r  \cdot \left( \mathbf{x} - \mathbf{X}_r(t) \right)} \times \notag \\
    & \times\, \psi_\mathrm{c}^{(r)}(\mathbf{x}-\mathbf{X}_r(t),t) \,,
\end{align}
where the phase $\mathcal{S}_{rs} / \hbar$ between the initial state $s$ and final state $r$ can be interpreted in terms of classical actions and decomposed as
 \begin{align}
  \label{eq:classicalaction}
      \mathcal{S}_{rs} &= \mathcal{S}_{s} (t_j-t_\mathrm{i})+\mathcal{S}_{r} (t_\mathrm{f}-t_j) \,,
  \end{align}
where $t_\mathrm{i}$ and $t_\mathrm{f}$ are the start and end times of the pulse, and $t_j$ is a arbitrarily chosen reference time corresponding to a hypothetical delta pulse. This decomposition models the evolution as free motion under the classical action $\mathcal{S}_s$ up to $t_j$, followed by evolution under $\mathcal{S}_r$ thereafter, yielding mnemonic trajectories that follow distinct free-evolution paths depending on the central momenta. The individual action contributions depend on the central momentum $\mathbf{P}_s$ and internal energy $E_s$ of the respective states:
   \begin{align}
      \label{eq:classicalaction2}
      \mathcal{S}_{s} &= \left(\mathbf{P}_s^2/(2m) - E_s \right) (t_k-t_l) \,. 
  \end{align}
This structure reflects the effective sequence of delta pulses separated by free evolution, supplemented by correction terms arising from the finite pulse. The list of $\mathbf{p}'$-independent phase contributions is summarized in Table\,\ref{tab:phaseterms}. These include $\phi^{(j)}_{rs}$, arising from the rotation back to the Schr\"odinger picture, and $\phi^{(\mathrm{sq})}_{rs}$, originating from the finite-pulse propagator $G_{rs}^\mathrm{(rot)}$.\\
\\
The time-dependent Hamiltonian contains oscillatory terms that can be eliminated by a unitary transformation to the so-called co-rotating frame, introducing the phase and frequency terms
\begin{equation}
  \label{eq:phij}
    \varphi_\pm^{(j)} = \varphi_{\pm} (t_0) + \Delta \omega_\pm^{(j)} (t_j-t_0) \,,
\end{equation}
and
\begin{equation}
\label{eq:omegaj}
    \Delta  \omega_\pm^{(j)} =  \Delta \omega_{\pm} (t_0) \,,
\end{equation}
which holds in the absence of residual acceleration or uncompensated frequency chirp. In the presence of an uncompensated chirp, this expression is replaced by the time-dependent definition given in Eqs.~\eqref{eq:omegacorot} and~\eqref{eq:omegaknotdefinition}, as detailed in Sec.~\ref{sec:chirpgeneral} (see also Appendix~\ref{app:chirp}).\\
\\
When transforming back to the Schr\"odinger picture, an additional phase term $\phi^{(j)}_{rs}$ arises, which is proportional to the detuning Eq.\,\eqref{eq:detuning}. This phase term is listed in the first column of Table\,\ref{tab:phaseterms}, with derivation provided in Appendix\,\ref{app:COM}.\\
\\
In the co-rotating frame, the Hamiltonian is given by
\begin{align}
\label{eq:rotatingHamiltonian}
    \hat{H}^\mathrm{(rot)} &= \sqrt{2} \Omega(t) \hat{J}_1 + \delta_\mathrm{A} \hat{J}_3 + \hbar \delta_\mathrm{S} \frac{\hat{J}_3^2}{\hbar^2} \notag \\
    &\approx \Omega_\mathrm{D}(t) \hat{J}_1 + \delta_\mathrm{A} \hat{J}_3 \,,
\end{align}
 where $\Omega_\mathrm{D}=\sqrt{2}\Omega$ is the generalized Rabi frequency of the three-level system, with the factor $\sqrt{2}$ arising from the normalization of the generator $\hat{J}_1$.
 Under appropriate conditions discussed in Sec.~\ref{sec:AC-Stark-shift}, the symmetric detuning $\delta_\mathrm{S}$ does not contribute to the total interferometric phase shift, and is neglected in the present treatment but may be included perturbatively in future extensions (for details see Appendix\,\ref{app:ALthreelevel}). The contribution involving the symmetric detuning $\delta_\mathrm{S}$ and $\hat{J}_3^2$ is not expressible as a linear combination of SU(2) generators, and therefore extends beyond the SU(2) algebraic framework.\\
\\
The generalized Euler formula is applied to obtain the propagator $ G_{rs}^\mathrm{(rot)} \big( \mathbf{p}', t, t_\mathrm{i} \big)$, as detailed in Appendix\,\ref{app:Grot}. The evolution of the wave packet in position space is obtained via the inverse Fourier transform
\begin{widetext}
\begin{align}
    e^{i \phi^{(\mathrm{sq})}_{rs}} \psi_\mathrm{c}^{(r)}(\mathbf{x}-\mathbf{X}_r(t),t) &=  \frac{1}{(2 \pi \hbar)^{3/2}} \, \int d^3p'  \, G_{rs}^\mathrm{(rot)} \big( \mathbf{p}', t, t_\mathrm{i} \big) \, e^{-i \frac{\mathbf{p}'^2 t}{2m\hbar}} \,e^{\frac{i}{\hbar}\mathbf{p}'  \cdot \left( \mathbf{x} - \mathbf{X}_s(t) \right)} \,\tilde{\psi}^{(s)}_\mathrm{c}(\mathbf{p}',t_\mathrm{i})
    \label{eq:Grot} ,
\end{align}
\end{widetext}
 where the propagator $G_{rs}^\mathrm{(rot)}$ is defined in Eq.\,\eqref{eq:Grotdefinition} and depends on the shape of the pulse envelope (here, a square pulse). Moreover, the phase correction $\phi^{(\mathrm{sq})}_{rs}$ has been explicitly factored out on the left-hand side of Eq.~\eqref{eq:Grot}. The matrix elements of the unitary evolution operator are proportional to the propagator and a Dirac delta function $\delta^{(3)}(\mathbf{p}'_f - \mathbf{p}'_i)$ (Eq.\,\eqref{eq:deltainmomentum}). As a result, the relative momentum $\mathbf{p}'$ with respect to the central momentum is preserved throughout the process: $\mathbf{p}'_i = \mathbf{p}'_f = \mathbf{p}'$.\\
 \\
 The $\mathbf{p}'$-dependent phase terms are categorized as:
  \begin{enumerate}[(a)]
    \item Linear: arising from the initial wave packet and the Fourier transform, as well as additional contributions from $G_{rs}^\mathrm{(rot)}$ that shift the central trajectory from $\mathbf{X}_s$ to $\mathbf{X}_r$,
    \item Quadratic: describe the expansion of the centered wave packet as if it were freely evolving,
    \item Higher-order: induces wave packet distortions (also from $\mathbf{p}'$-dependent modulus), as discussed in \cite{Augst2027}.
  \end{enumerate}
  The structuring of the $\mathbf{p}'$-independent phase terms to account for shifts in the central trajectory naturally leads to a transition from the Hamiltonian to the Lagrangian framework, the emergence of the classical action in the phase, the sytematic organization in terms of mnemonic trajectories (dashed lines in the Figures of this paper) with a simple intuitive interpretation, as well as correction phases $\phi^{(\mathrm{sq})}_{rs}$ due to the propagator $G_{rs}^\mathrm{(rot)}$, listed in the second column of Table\,\ref{tab:phaseterms}.
\begin{figure*}[t!]
    \centering
    \begin{subfigure}[t]{0.32\textwidth}
        \centering
        \caption{}
        \vspace*{.25cm}
        \hspace*{-.75cm} \includegraphics[width=1\textwidth]{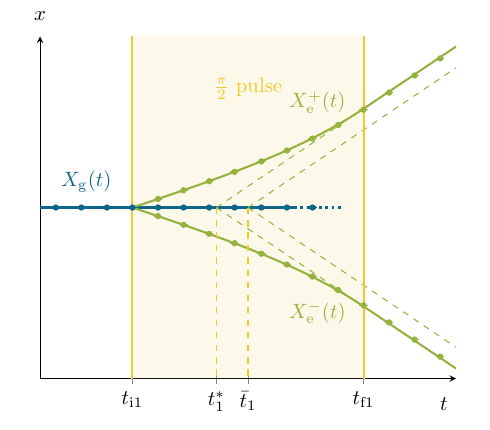}
        \label{fig:double_raman_bs}
    \end{subfigure}
    \hfill
    \begin{subfigure}[t]{0.32\textwidth}
        \centering
        \caption{}
       \hspace*{-.75cm} \includegraphics[width=1\textwidth]{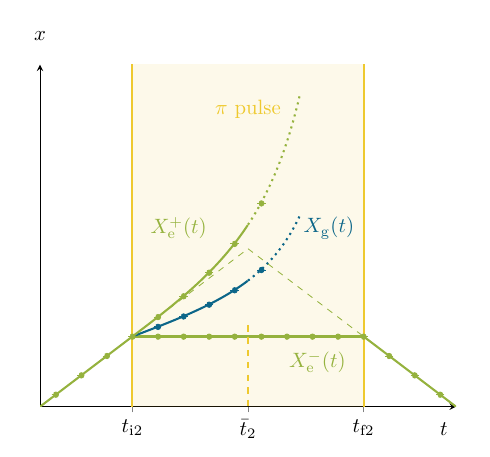}
        \label{fig:mirror1}
    \end{subfigure}
    \hfill
    \begin{subfigure}[t]{0.32\textwidth}
        \centering
        \caption{}
       \hspace*{-1cm}  \includegraphics[width=1\textwidth]{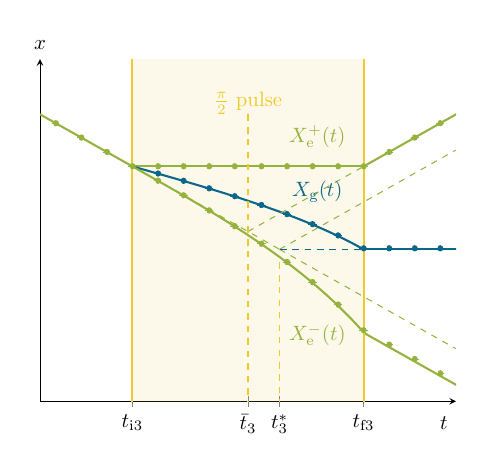}
        \label{fig:bs_upper}
    \end{subfigure}
    \caption{Step-by-step analysis of each pulse in the upper branch of a Mach-Zehnder interferometer: the first $\pi/2$ beam-splitter pulse, starting from the ground state (a), the mirror ($\pi$) pulse, starting from the upper excited state (b), and the final beam-splitter pulse, starting from the lower excited state (c), as described in Secs.\,\ref{sec:firstBS} and \ref{sec:mirrorandsecondBS}. Analytical trajectories (solid lines) are compared with numerical results (markers) obtained using the simulation approach of Ref.\,\cite{Augst2027}, using an initial Gaussian wave packet with \(\sigma_\mathrm{p} = 0.02 \,\hbar \mathbf{k}_\mathrm{eff}\) and pulse duration \(\tau = 100~\mu\text{s}\). The solid curves show the exact analytical predictions for the central trajectories $X_r(t)$. Dotted lines indicate regimes where the centered-wave-packet approximation breaks down, while dashed lines show the mnemonic trajectories, using $\bar{t}_j$ at the pulse center and, for the beam-splitter pulses, $t_j^*$. (a) The initial ground-state component (blue) is coherently split into upper and lower excited-state components (green), while the undiffracted ground-state amplitude is strongly suppressed. (b,c) The dominant doubly diffracted component (green) remains stationary, while residual undiffracted excited-state (green) and singly diffracted ground-state (blue) components exhibit time-dependent deflections.}
    \label{fig:double_raman_schematic}
\end{figure*}
\begin{table*}[]

    \centering
    \begin{tabular}{c  l l }
        \toprule
        & $\phi^{(j)}_{rs}$ connected to laser phase at $t_j$
         & 
        $\phi^{(\mathrm{sq})}_{rs}$ correction phase for square pulse  \\
        \midrule
        
        \begin{tikzpicture}[scale=0.5,baseline=0ex]
            \draw[dlrblue,line width=1pt] (-0.8,0)--(0.8,0);
            \draw[dlryellow,line width=1pt] (0,-0.8)--(0,0.8);
        \end{tikzpicture}
        &
        $\phi^{(j)}_\mathrm{gg}=0$
        &
        $\phi^{(\mathrm{sq})}_\mathrm{gg}=0$
        \\
        
        \begin{tikzpicture}[scale=0.5,baseline=0ex]
            \draw[dlrgreen,line width=1pt] (-0.8,-0.8)--(0,0);
            \draw[dlryellow,line width=1pt] (0,-0.8)--(0,0.8);
            \draw[dlrgreen,line width=1pt] (0,0)--(0.8,0.8);
        \end{tikzpicture}
        &
        $\phi^{(j)}_\mathrm{\pm \pm}
        =
        \delta_\pm^{(\mathrm{c})}(t_\mathrm{f}-t_\mathrm{i})$
        &
        $\begin{aligned}
        \phi^{(\mathrm{sq})}_\mathrm{\pm \pm}
        ={}&
        \mp
        \delta_\mathrm{A}^{(\mathrm{c})} 
        2 \Delta X(t_\mathrm{f}-t_\mathrm{i})/v_{\mathrm{rec}}
        \end{aligned}$
        \\
        
        \begin{tikzpicture}[scale=0.5,baseline=0ex]
            \draw[dlrgreen,line width=1pt] (-0.8,-0.8)--(0,0);
            \draw[dlryellow,line width=1pt] (0,-0.8)--(0,0.8);
            \draw[dlrblue,line width=1pt] (0,0)--(0.8,0);
        \end{tikzpicture}
        &
        $\phi^{(j)}_\mathrm{g \pm}
        =
        \delta_\pm^{(\mathrm{c})}(t_j-t_\mathrm{i})
        +\varphi_\pm^{(j)}
        \mp
        \mathbf{k}_\mathrm{eff}\cdot\mathbf{X}^{(j)}_\pm$
        &
        $\begin{aligned}
        \phi^{(\mathrm{sq})}_\mathrm{g \pm}
        ={}&
        \mp
        \delta_\mathrm{A}^{(\mathrm{c})}
        \Delta X(t_\mathrm{f}-t_\mathrm{i})/v_{\mathrm{rec}}
        \end{aligned}$
        \\
        
        \begin{tikzpicture}[scale=0.5,baseline=0ex]
            \draw[dlrblue,line width=1pt] (-0.8,0)--(0,0);
            \draw[dlryellow,line width=1pt] (0,-0.8)--(0,0.8);
            \draw[dlrgreen,line width=1pt] (0,0)--(0.8,0.8);
        \end{tikzpicture}
        &
        $\phi^{(j)}_\mathrm{\pm g}
        =
        \delta_\pm^{(\mathrm{c})}(t_\mathrm{f}-t_j)
        -
        \varphi_\pm^{(j)}
        \pm
        \mathbf{k}_\mathrm{eff}\cdot\mathbf{X}^{(j)}_\mathrm{g}$
        &
        $\begin{aligned}
        \phi^{(\mathrm{sq})}_\mathrm{\pm g}
        =&
        \mp \delta_\mathrm{A}^{(\mathrm{c})}
        \Delta X(t_\mathrm{f}-t_\mathrm{i})/v_{\mathrm{rec}}
        \end{aligned}$
        \\
        
        \begin{tikzpicture}[scale=0.5,baseline=0ex]
            \draw[dlrgreen,line width=1pt] (-0.8,0.8)--(0,0);
            \draw[dlrgreen,line width=1pt] (0,0)--(0.8,0.8);
            \draw[dlryellow,line width=1pt] (0,-0.8)--(0,0.8);
        \end{tikzpicture}
        &
        $\begin{aligned}
        \phi^{(j)}_\mathrm{\pm \mp}
        ={}&
        \delta_\pm^{(\mathrm{c})}(t_\mathrm{f}-t_j)
        +
        \delta_\mp^{(\mathrm{c})}(t_j-t_\mathrm{i})
        \mp\varphi_+^{(j)}
        \pm\varphi_-^{(j)}
        \pm2\mathbf{k}_\mathrm{eff}\cdot\mathbf{X}^{(j)}_\mp
        \end{aligned}$
        &
        $\phi^{(\mathrm{sq})}_\mathrm{\pm \mp}
        = 0$
        \\
        
        \bottomrule
        \end{tabular}
    \caption{Spatially-independent phase contributions ($\mathbf{p}'$-independent in momentum representation) systematically organized in terms of menmonic trajectories: $\phi^{(j)}_{rs}$, arising from the rotation back to the Schr\"odinger picture, and $\phi^{(\mathrm{sq})}_{rs}$, originating from the propagator $G_{rs}^\mathrm{(rot)}$ connected to the finite-pulse evolution in the co-rotating frames (this contribution depends on the shape of the pulse envelope and is given here for the case of a square pulse). The free-evolution segments in the ground state (blue) and the upper/lower excited states (green) are characterized by the classical actions $\mathcal{S}_\mathrm{g}$ and $\mathcal{S}_\mathrm{\pm}$, respectively. The laser pulse (yellow) at an arbitrarily chosen reference time $t_j$ partitions the interferometer sequence into two propagation intervals, with the actions evaluated over the corresponding time domains (e.g., $\mathcal{S}_{+}(t_j -t_\mathrm{i})$ before and $\mathcal{S}_{-}(t_\mathrm{f} -t_j)$ after the pulse). The central positions $\mathbf{X}^{(j)}_s$ at time $t_j$ in the co-moving frame are defined in Appendix\,\ref{app:COM}. Additional phase terms in $\phi^{(\mathrm{sq})}_{rs}$ that are proportional to $\mathbf{P}_r\cdot\Delta\mathbf{X}_{rs}$ (Sec.\,\ref{sec:tj}) have been omitted, as they cancel out between the two interferometer arms and do not contribute to the measurable phase shift.}
    \label{tab:phaseterms}
\end{table*}
\subsection{First beam-splitter pulse}
\label{sec:firstBS}
The phase terms linear in $\mathbf{p}'$ induce a shift in the central trajectory of the atomic wave packet. For the undiffracted ground state, the trajectory remains identical to that of an ideal delta pulse:
  \begin{equation}
  \label{eq:trajXgg}
      \mathbf{X}_\mathrm{g}(t) = \mathbf{X}_\mathrm{i}+\frac{\mathbf{P}_\mathrm{g}}{m_\mathrm{g}} \cdot (t - t_\mathrm{i}) \,,
  \end{equation}
    with the initial position $\mathbf{X} (t_\mathrm{i}) \equiv \mathbf{X}_\mathrm{i}$. 
    (Note that in this section we explicitly include a non-vanishing central momentum $\mathbf{P}_\mathrm{g}$ for the initial wave packet, but in the following sections, where the freely falling frame co-moving with the mid-point trajectory will be considered, one has $\mathbf{P}_\mathrm{g} = \mathbf{0}$.)
    The trajectory $\mathbf{X}_\mathrm{g}(t)$, along with all subsequently presented trajectories, is valid during the pulse, i.e., for \( t_\mathrm{i} < t < t_\mathrm{f} \). After the pulse, the trajectories reduce to those obtained for a delta pulse, with the addition of a shift in phase and position accumulated during the pulse.\\
    \\
  In contrast, the diffracted excited states follow a modified central trajectory:
  \begin{equation}
  \label{eq:trajXgepm}
      \mathbf{X}^\pm_\mathrm{e}(t) = \mathbf{X}_\mathrm{i}+\frac{\mathbf{P}_\mathrm{g}}{m} \cdot(t - t_\mathrm{i}) \pm \Delta \mathbf{X}(t-t_\mathrm{i}) \,,
  \end{equation}
  where the signs ($+$) and ($-$) correspond to the positive- and negative-recoil  branches, respectively.
  The trajectory shift $\Delta \mathbf{X}$, which arises from the additional phase contributions contained within the propagator $G_{rs}^\mathrm{(rot)}$, is defined as the following vector-valued time-dependent function:
  \begin{equation}
  \label{eq:deltaX}
      \Delta \mathbf{X}(t-t_\mathrm{i}) \equiv \frac{\mathbf{v}_\mathrm{rec}}{\Omega_\mathrm{D} } \tan \left(\frac{\Omega_\mathrm{D} (t - t_\mathrm{i})}{2}\right) \,.
  \end{equation}
  This result assumes narrow wave packets ($\delta' \ll \Omega$) for which the effective Rabi frequency (Eq.~\eqref{eq:omegaeff}) can be approximated by
  \begin{equation}
  \label{eq:approxomegaD}
      \Omega_\mathrm{eff} \approx \Omega_\mathrm{D} \,.
  \end{equation}
  Figure\,\ref{fig:double_raman_bs} shows the central trajectories during the first beam-splitter pulse, initially in ground state, with the diffracted excited states displaced symmetrically by $\Delta \mathbf{X}(t-t_\mathrm{i})$. The trajectory shift derived from the approximation in Eq.~\eqref{eq:propagatorapprox} holds for small detunings and when the wave packet shape remains unchanged. This condition is satisfied for most of the pulse area, except for the residual component near the mirror pulse (for the residual undiffracted component) and near the $2\pi$ pulse (for the residual diffracted component). In these regimes, the residual wave packet shape deviates significantly, and the approximation breaks down, as detailed in Ref.~\cite{Augst2027}.
\subsection{Mirror and second beam-splitter pulses}
\label{sec:mirrorandsecondBS}
The central trajectory of the undiffracted excited state $\mathbf{X}^\pm_\mathrm{e}(t)$, is given by
  \begin{equation}
    \label{eq:trajfullyidffracted}
      \mathbf{X}^\pm_\mathrm{e}(t) \approx \mathbf{X}_\mathrm{i} + \frac{\mathbf{P}_\mathrm{g}}{m}\cdot(t - t_\mathrm{i}) \pm 2\Delta \mathbf{X}(t - t_\mathrm{i}) \,,
  \end{equation}
  where 
  the factor of 2 arises from approximating the effective Rabi frequency in Eq.\,\eqref{eq:approxomegaD} by the generalized three-level Rabi frequency $\Omega_\mathrm{D} = \sqrt{2}\Omega$.\\
  \\
  The diffracted ground-state component evolves according to
  \begin{equation}
  \label{eq:trajXepmtog}
      \mathbf{X}_\mathrm{g}(t) = \mathbf{X}_\mathrm{i} + \frac{\mathbf{P}_\mathrm{g}}{m_\mathrm{g}}\cdot(t - t_\mathrm{i}) + \Delta \mathbf{X}(t - t_\mathrm{i}) \,.
  \end{equation}
  On the other hand, The diffracted excited-state component follows a static trajectory in the co-moving frame:
  \begin{equation}
  \label{eq:trajXdiffe}
      \mathbf{X}^\mp_\mathrm{e}(t) = \mathbf{X}_\mathrm{i} + \frac{\mathbf{P}_\mathrm{g}}{m}\cdot(t - t_\mathrm{i}) \,.
  \end{equation}
 \\
Figures~\ref{fig:mirror1} and~\ref{fig:bs_upper} illustrate the central trajectories during the mirror pulse and the second beam-splitter pulse of the upper arm, starting from the excited state in the positive- and negative-recoil branches, respectively. For the mirror pulse, the trajectory $\mathbf{X}^\mp_\mathrm{e}(t)$ given in Eq.~\eqref{eq:trajXdiffe} dominates, while the amplitudes of the two residual components vanish in the limit of $\mathcal{A}\to \pi$. In the final beam-splitter pulse, all three trajectories, leading to the three output ports, are relevant and contribute to the interference.
\subsection{Convenient choice for $t_j$}
\label{sec:tj}
 Transforming back to the Schr\"odinger picture introduces phase terms $\varphi_\pm ^{(j)}$ and terms that depend on the frequency difference $\Delta\omega_\pm ^{(j)}$ between counter-propagating lasers (Eqs.~\eqref{eq:phij} and~\eqref{eq:omegaj}). Combined with the internal-state energies, the latter can be expressed in terms of the detuning $\delta_\pm$ defined in Eq.~\eqref{eq:detuning} (see left column of Table\,\ref{tab:phaseterms}; details in Appendix\,\ref{app:COM}). In accordance with the momentum decomposition in Eq.~\eqref{eq:momentumdecomposition}, the detuning is split into a central component $\delta^{(\mathrm{c})}_\pm$ and a term depending on the relative momentum $\mathbf{p}'$ as
  \begin{equation}
  \label{eq:detuningdecomposed}
      \delta_\pm = \delta^{(\mathrm{c})}_\pm + \frac{\mathbf{p}' \cdot \mathbf{k}_\mathrm{eff}}{m} \,.
  \end{equation}
  In the phase terms associated with $G_\mathrm{rot}$ (second column of Table\,\ref{tab:phaseterms}), the terms involving the symmetric detuning $\delta_\mathrm{S}$ have been omitted because, under the conditions discussed in Sec.~\ref{sec:AC-Stark-shift}, their contributions to the total interferometer phase shift cancel out, leaving only the antisymmetric component $\delta_\mathrm{A}$ relevant for the measurable phase shift.\\
  \\
  The apparent $t_j$-dependence of the phase terms $\phi^{(j)}_{rs}$ in Table\,\ref{tab:phaseterms} cancels out upon full evaluation (see Sec.\,\ref{sec:chirpgeneral} and Appendix\,\ref{app:ALthreelevel}). The parameter $t_j$, which can be freely chosen, defines the mnemonic trajectories (dashed lines in Fig.~\ref{fig:double_raman_schematic}). As discussed in Appendix~\ref{app:Grot}, adjusting the plane-wave factor in Eq.~\eqref{eq:psigeneral} to match the actual central trajectories introduces a phase shift proportional to $\mathbf{P}_r \cdot \Delta\mathbf{X}_{rs}/\hbar$, where $\Delta\mathbf{X}_{rs}$ (defined in  Eq.\,\eqref{eq:omittedspatiallydepphaseterms}) quantifies the difference between the mnemonic and actual trajectories. Although the choice of $t_j$ typically results in $\Delta\mathbf{X}_{rs} \neq 0$, in the freely falling frame comoving with the mid-point trajectory this phase contribution is identical in both arms of the interferometer and therefore cancels out in phase shift measurements.\\
 \\
  There is a specific choice of times $t_j^*$ for which the mnemonic trajectories coincide at the end of the pulse with the actual central trajectories  ($\Delta\mathbf{X}_{rs} = 0$) and corresponds to
  \begin{align}
  \label{eq:tstar}
      t_1^* - t_\mathrm{i1} &= \tau_1 \left( 1- \frac{1}{\Omega_\mathrm{D}^{(1)} \tau_1} \tan \left(\frac{\Omega_\mathrm{D}^{(1)} \tau_1}{2}\right)\right) \notag \\
      t_3^* - t_\mathrm{i3} &= \tau_3 \frac{1}{\Omega_\mathrm{D}^{(3)} \tau_3} \tan \left(\frac{\Omega_\mathrm{D}^{(3)} \tau_3}{2}\right) \,,
  \end{align}
  where the superscripts $1,3$ label the first and third beam-splitter pulses, with pulse durations $\tau_1 = t_\mathrm{f1} - t_\mathrm{i1}$ and $\tau_3 = t_\mathrm{f3} - t_\mathrm{i3}$, respectively. Note that to obtain $\Delta\mathbf{X}_{rs} = 0$ for the main component of the mirror pulse, the reference time $t_2$ is chosen at the pulse center, denoted by $\bar{t}_2$.\\
    \\
  A common convention in earlier treatments is to choose the pulse center as the reference time \cite{Antoine_2006}, often referred to as the ``time origin''. The derived times $t_j^*$ share a formal similarity with the time origin introduced in Ref.~\cite{Saywell2026}, though their physical interpretations differ. In Ref.~\cite{Saywell2026},
  the redefined time origin is introduced to get a simple scaling $T^2$ for the response to accelerations in single-diffraction interferometry when terms quadratic in the pulse duration ($\tau^2$) are neglected. In contrast, our $t_j^*$ is defined for double diffraction and corresponds to the choice of $t_j$ for which the mnemonic trajectories and the actual central trajectories coincide both at the beginning and at the end of the pulse. Moreover, we explicitly include terms quadratic in $\tau$.

\section{Full atom interferometer}
\label{sec:fullAI}
\subsection{Central trajectories: closed and open interferometers}
\begin{figure*}[t!]
    \centering
    \begin{subfigure}[t]{0.6\textwidth}
        \hfill
        \caption{}
        \includegraphics[width=1.1\textwidth]{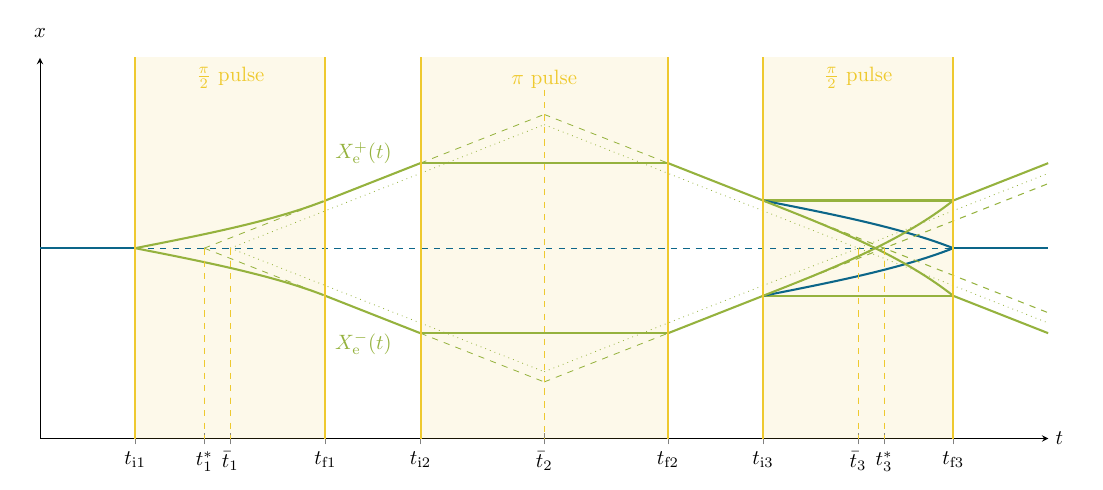}
        \label{fig:fullAI}
    \end{subfigure}
    \hfill
    \begin{subfigure}[t]{0.3\textwidth}
        \caption{}
        \hspace{-1.5cm}
        \includegraphics[width=1.05\textwidth]{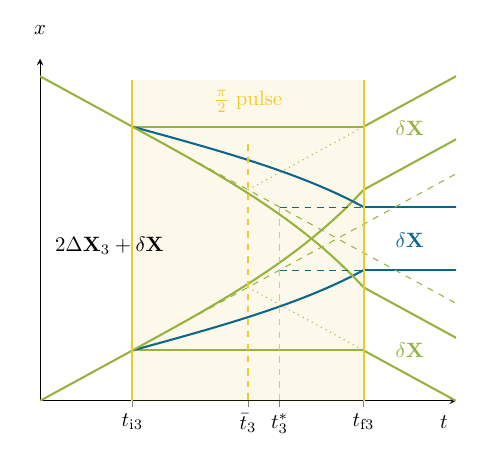}
        \captionsetup{width=1.3\textwidth, margin={-1.2cm, 0cm}}
        \label{fig:openAI}
    \end{subfigure}
    \caption{Central trajectories of the full atom interferometer for (a) the closed and (b) the open configuration. Wave-packet components are color-coded as in Fig.\,\ref{fig:double_raman_schematic}. (a) Closed interferometer configuration. Nontrivial trajectory shifts induced by the first and third beam-splitter pulses cancel out exactly in this configuration. The dashed lines indicate the mnemonic trajectories for the reference time $t_j^*$, at which the mnemonic path coincides with the actual trajectory at the end of the pulse, while the dotted lines correspond to the choice of symmetric reference time $\bar{t}_j$. (b) Open interferometer configuration. An additional trajectory shift $\delta \mathbf{X}$, arising from imbalances in pulse durations, effective Rabi frequencies, interrogation times, pulse areas, or the presence of gravity gradients, leads to a spatial separation of the wave-packet components at all output ports.}
    \label{fig:double_raman_full}
\end{figure*}

A closed interferometer requires that the trajectory shift induced by the third pulse, $\Delta \mathbf{X}_3 \equiv \Delta \mathbf{X}(\tau_3)$, exactly compensates the arm separation at the beginning of the pulse. For the central exit port this condition can be easily understood by considering the time-reversal of the initial beam-splitter pulse. As shown in Figure~\ref{fig:fullAI}, a shift of $-\Delta \mathbf{X}_3$ in the upper arm and $+\Delta \mathbf{X}_3$ in the lower arm (see Sec.~\ref{sec:single_pulse}) results in perfect overlap at the central output port after the final beam splitter. For the side ports, the double-diffracted component remains unshifted, while the undiffracted component shifts by $2\Delta \mathbf{X}_3$. In all cases, the interferometer remains closed. \\
\\
An open interferometer arises either through pulse parameter imbalances ($\Omega_1 \neq \Omega_3$, $\tau_1 \neq \tau_3$, or $T_1 \neq T_2$) or in the presence of gravity gradients, even for nominally balanced sequences. To quantify this, consider a small deviation $\delta \mathbf{X}$ from the closed condition, i.e., $\Delta \mathbf{X}_3 \to \Delta \mathbf{X}_3 + \delta \mathbf{X}$, as depicted in Figure~\ref{fig:openAI}. By symmetry, the trajectories contributing to the central output port are shifted by $\mp \delta \mathbf{X}/2$. For the side ports, the fully diffracted component remains unshifted, while the undiffracted component shifts by $\pm \delta \mathbf{X}$ for the upper and lower ports, respectively. As a result, the interferometer is effectively open by a displacement $\delta \mathbf{X}$ in all output ports.
\subsection{Phase-shift contributions}
 We focus on the case of time-independent detuning, corresponding to zero acceleration and the absence of frequency chirp (chirp effects are addressed in the following section). In the co-moving frame, the symmetry of double diffraction ensures that many standard phase contributions are identical for both interferometer arms and thus cancel out in the phase shift. We first analyze the phase contributions at the central exit port, generalizing to the side ports subsequently. 
 \subsubsection{Phase-shift contributions at the central exit port}
 The total phase arises from three distinct origins:
 \\
 \paragraph{Propagation phases:}
 This includes both true free evolution (including effects of gravity gradients) and the propagation during the pulse, which is decomposed into two segments: a mnemonic free-evolution segment up to the reference time $t_j$ (with initial momentum) and a second segment from $t_j$ onward (with final momentum). The mnemonic free-evolution contribution is governed by the classical action in Eq.\,\eqref{eq:classicalaction2}. In symmetric configurations, these free-evolution phases are identical for both arms and cancel out in the differential signal. In addition to this, the total phase includes extra contributions: the term $\phi^{(j)}_{rs}$ from the left column of Table~\ref{tab:phaseterms}, and additional spatial correction proportional to $\mathbf{P}_r \cdot \Delta\mathbf{X}_{rs}/\hbar$, as detailed in Appendix~\ref{app:Grot}, specifically in Eq.~\eqref{eq:omittedspatiallydepphaseterms}. For the central output port, the phase shifts are proportional to $ \pm \mp \mathbf{k}_\mathrm{eff} \cdot \delta \mathbf{X}/2$ from the upper and lower arms, respectively. These contributions also cancel out in the interferometric phase due to symmetry.
 \\
 \paragraph{Separation phase:}
 Defined as $\phi_\mathrm{sep} = -\mathbf{P} \cdot \delta\mathbf{X}/\hbar$ (Eq.\,(C10) in Ref.\,\cite{Roura2020}), this term vanishes for a closed interferometer. At the central port in the co-moving frame, $\mathbf{P}_\mathrm{g}=0$, so the separation phase is zero in any case.
 \\
 \paragraph{Phases from laser pulses:} In the co-moving frame, the spatially dependent contribution vanishes due to symmetry. The spatially independent part decomposes into a delta-pulse-like laser phase contribution $\delta\phi_\mathrm{laser}$ and a contribution $\delta\phi_\mathrm{det}$ proportional to the detuning describing the dynamics during the pulse:
   \begin{equation}
      \label{eq:fullAIphasediff}
          \delta \phi = \delta \phi_\mathrm{laser} + \delta \phi_\mathrm{det} \,.
      \end{equation}
\textbf{Laser phases $\delta \phi_\mathrm{laser}$:} 
   The delta-pulse-like laser phase contribution reads
      \begin{align}
      \label{eq:laserphasecontri}
          \delta \phi_\mathrm{laser} &= -\varphi^{(1)}_+ +\varphi^{(1)}_-  + 2 \varphi^{(2)}_+ - 2\varphi^{(2)}_-  - \varphi^{(3)}_+ + \varphi^{(3)}_- \notag \\
          &= - 2 \Delta\omega_\mathrm{A} (t_1 - 2 t_2 + t_3)  + \delta \varphi_0 \,,
      \end{align}
      where $\varphi^{(j)}_{\pm}$ are defined in Eq.\,\eqref{eq:phij}, and we have introduced $\delta\phi_0$, defined as 
      \begin{align}
        \label{eq:initiallaserphase}
        \delta \varphi_0 &= -\varphi^{(1)}_+ (t_0) +\varphi^{(1)}_- (t_0) +2\varphi^{(2)}_+ (t_0) -2\varphi^{(2)}_- (t_0) \notag \\
          &-\varphi^{(3)}_+ (t_0) +\varphi^{(3)}_- (t_0) \,.
    \end{align}
    In laser pulses driving two-photon transitions, the two-photon frequencies $\Delta \omega_\pm (t)$, and the corresponding phase $\varphi_\pm (t)$, are stabilized by locking them to a radio-frequency or microwave oscillator. Therefore, in an ideal situation $\Delta\omega_\mathrm{A}(t_1)=\Delta\omega_\mathrm{A}(t_2)=\Delta\omega_\mathrm{A}(t_3)=\Delta\omega_\mathrm{A}$, as assumed in the second equality of Eq.\,\eqref{eq:laserphasecontri}, and $\delta\varphi_0 = 0$.\\
    \\
    However, imperfections in the reference oscillator or the locking process, and vibrations of the optical fibers in parts of the optical paths that are not common to both counter-propagating frequency components will lead in practice to laser frequency noise (and related laser phase noise). This can be taken into account by considering a time-dependent frequency chirp, as done in Ref.~\cite{Roura_2025}, and an explicit calculation would then involve the methods introduced in Secs.~\ref{sec:chirpgeneral} and \ref{sec:time-dependent_chirp}. Nevertheless, laser phase noise taking place between pulses can be effectively taken into account by changing the value of $\varphi_\pm^{(j)} (t_0)$ employed for each pulse, which leads in general to a non-vanishing $\delta\varphi_0$.\\
    \\
    For double diffraction with a retro-reflecting set-up in microgravity (or horizontal configurations), laser phase noise is the same for the two opposite pairs of counter-propagation beam, so that $\varphi_+^{(j)} (t_0) = \varphi_-^{(j)} (t_0)$ and $\delta\varphi_0 = 0$. On the other hand, in ground-based vertical configurations this is no longer the case.\\
    \\
    \textbf{Effects during the pulse $\delta \phi_\mathrm{det}$:} 
    The contribution of the dynamics during the pulse is given by
     \begin{align}
     \label{eq:phidet}
        \delta \phi_\mathrm{det} &= 2 \delta_\mathrm{A}^{(\mathrm{c})} (t_1) \, (t_\mathrm{f1}-t_1)
         - 2\delta_{\mathrm{A}}^{(\mathrm{c})}(t_3) \, (t_3-t_\mathrm{i3}) \notag \\
        &-  2 \frac{\delta_\mathrm{A}^{(\mathrm{c})} (t_1)}{ \Omega_\mathrm{D}^{(1)}  } \tan \left(\frac{\mathcal{A}_1}{2}\right) + 2 \frac{\delta_\mathrm{A}^{(\mathrm{c})} (t_3)}{ \Omega_\mathrm{D}^{(3)}  }  \tan \left(\frac{\mathcal{A}_3}{2} \right) \,,
      \end{align}
      with the pulse area of the $j$th square pulse given by \(\mathcal{A}_j = \Omega_\mathrm{D}^{(j)} \tau_j\).\\
      \\
      The detuning $\delta_\pm$ is defined in Eq.\,\eqref{eq:detuning}. In the phase shift, the symmetric component $\delta_\mathrm{S}$ vanishes if it is homogeneous across the arm separation. For the propagator $G_{rs}^\mathrm{(rot)}(\mathbf{p}', t, t_\mathrm{i})$, we set $\delta_\mathrm{S} \to 0$ from the outset. Generally, all symmetric contributions are identical for both arms at the central exit port and thus cancel out. For the mirror pulse, the antisymmetric detuning $\delta_\mathrm{A}$ contributes only if $t_2 \neq \bar{t}_2$; choosing $t_2 = \bar{t}_2$ (pulse center) eliminates this term. In a balanced interferometer, where $t_1 - 2t_2 + t_3 = 0$, $\delta_\mathrm{A}^{(\mathrm{c})}(t_1) = \delta_\mathrm{A}^{(\mathrm{c})}(t_3)$, $(t_\mathrm{f1}-t_1) = (t_3-t_\mathrm{i3})$, and $\mathcal{A}_1 = \mathcal{A}_3$ with $\Omega_\mathrm{D}^{(1)} = \Omega_\mathrm{D}^{(3)}$, all detuning-dependent terms cancel out.\\
      \\
    \subsubsection{Phase-shift contributions at the side exit ports}
    Extending the analysis to the side ports yields the same conclusion for the phase shift. 
    The spatially dependent terms follow a different cancellation mechanism: the separation phase proportional to $-\mathbf{k}_\mathrm{eff} \cdot \delta \mathbf{X}$ exactly cancels out the spatial correction term (Eq.~\eqref{eq:omittedspatiallydepphaseterms}). This cancellation occurs because the undiffracted component in each arm is shifted by $+2\mathbf{k}_\mathrm{eff} \cdot \delta \mathbf{X}/2$, while the fully diffracted component remains unshifted. As a result, the separation phase is fully compensated by the shift in the undiffracted path.\\
    \\
    Unlike at the central port, the symmetric detuning $\delta_\mathrm{S}$ does not automatically cancel out at the side ports, as the port selection breaks the spatial symmetry. However, this symmetry (and the resulting cancellation) can be restored by taking the semi-sum of the phase shifts obtained for both side ports. (Note that for $G_{rs}^\mathrm{(rot)}(\mathbf{p}', t, t_\mathrm{i})$, $\delta_\mathrm{S} \to 0$ is imposed from the start; see Sec.\,\ref{sec:single_pulse}.)
\subsection{Laser phases in different reference frames}
\label{subsec:fuallAIphase}
Transforming from the laboratory frame (where the lasers are at rest) to a freely falling frame introduces a time-dependent Doppler shift:
  \begin{equation}
      \Delta\omega_{\pm}'(t) = \Delta\omega_{\pm} (t) \mp \mathbf{k}_\mathrm{eff} \cdot \mathbf{v}(t) \,,
  \end{equation}
  with velocity
  \begin{equation}
      \mathbf{v}(t) = \mathbf{v} (t_0) + \int_{t_0}^{t} \mathbf{a}(t') \, dt' = \mathbf{v}(t_0) + \mathbf{a}\cdot(t - t_0) 
      \label{eq:vel},
  \end{equation}
   for constant acceleration $\mathbf{a}$. In vertical ground-based experiments, $\mathbf{a} \to \mathbf{g}$, and a laser frequency chirp $\alpha_\pm$ is applied to compensate for the time-dependent component:
  \begin{align}
      \Delta \omega_\pm^\mathrm{(L)} (t) &= \Delta\omega_\pm^\mathrm{(L)}(t_0) + \int_{t_0}^{t} \alpha_{\pm}(t') \, dt' \notag \\ &= \Delta\omega_\pm^\mathrm{(L)}(t_0) + \alpha_{\pm} \cdot (t - t_0) 
      \label{eq:omegawithchirp1},
  \end{align}
  where we use the superscripts (L) and (C) to denote quantities in the laboratory and co-moving frames, respectively. As shown in the second equality, a constant chirp rate $\alpha_\pm$ is assumed; the generalization to a time-dependent $\alpha_\pm(t)$ will be discussed in Sec.~\ref{sec:time-dependent_chirp}.\\
  \\
  In the freely falling frame, the laser frequency becomes
  \begin{align}
      \label{eq:omegacorotL}
    \Delta \omega_\pm^\mathrm{(C)} (t) 
    &= \Delta\omega_\pm^\mathrm{(L)} (t_0) \mp \mathbf{k}_\mathrm{eff}\cdot \mathbf{v}(t_0) + \Delta \alpha_\pm \cdot (t - t_0) \,,
  \end{align}
where the uncompensated frequency chirp is defined as
  \begin{equation}
      \Delta \alpha_\pm = \alpha_{\pm} \mp \mathbf{k}_\mathrm{eff}\cdot\mathbf{a} \,.
  \end{equation}
  By uncompensated frequency chirp we mean the residual chirp rate $\Delta \alpha_\pm$ that remains when imperfectly compensating the Doppler shift $\mathbf{k}_\mathrm{eff}\cdot\mathbf{a}$. Integrating the frequency relation $\omega = d\varphi/dt$ from $t_0$ to $t$ yields the phase in the freely falling frame:
  \begin{align}
  \label{eq:phicorotL}
        \varphi_\pm^\mathrm{(C)}(t) &= \varphi_{\pm}^\mathrm{(L)}(t_0) + \left(\Delta \omega^\mathrm{(L)}_{\pm}(t_0) \mp \mathbf{k}_\mathrm{eff}\cdot \mathbf{v}(t_0) \right) (t-t_0) \notag \\
        &+ \frac{1}{2} \Delta \alpha_{\pm} \cdot(t-t_0)^2 \,.
  \end{align}
  Eqs.\,\eqref{eq:omegacorotL} and\,\eqref{eq:phicorotL} are essential for characterizing the effects of uncompensated frequency chirp in atom interferometry in the following section. Note that the symmetric chirp $\alpha_\mathrm{S}$, defined via the decomposition into symmetric and antisymmetric components analogous to the detuning in Eqs.~\eqref{eq:detuningsymm} and~\eqref{eq:detuningantisymm}, receives no Doppler shift and its contributions to the phase shift cancel out in symmetric interferometers, by an argument parallel to that for the symmetric detuning. Consequently, only the antisymmetric component $\alpha_\mathrm{A}$ contributes, leading to the effective uncompensated chirp $\Delta\alpha_\mathrm{A} = \alpha_\mathrm{A} - \mathbf{k}_\mathrm{eff} \cdot \mathbf{a}$.

\section{Accelerations and frequency chirps}
\label{sec:chirpgeneral}
In the freely falling frame, the effects of accelerations and any frequency chirp applied to the lasers are entirely captured by $\Delta \alpha_\pm$. This includes the effects of residual accelerations in microgravity experiments as well as the effects of gravitational acceleration and accelerated laser frames that are not fully compensated by the frequency chirp in ground-based experiments.
For non-vanishing $\Delta \alpha_\pm$, the argument of the time-dependent phase factor appearing in the non-diagonal matrix elements of the Hamiltonian $H_\mathrm{int}$ becomes
\begin{align}
\varphi_\pm (t)= \varphi_\pm^{(j)} + \Delta \omega_\pm^{(j)}\, (t - t_j) + \frac{1}{2} \Delta \alpha_\pm \, (t - t_j)^2
\label{eq:phase-chirp} ,
\end{align}
which has been written in terms of the relative time variable $(t - t_j)$ for each pulse, and where $\varphi_\pm^{(j)}$ and $\Delta \omega_\pm^{(j)}$ correspond to the quantities evaluated at time $t_j$. The two-photon frequency difference $\Delta \omega_\pm^{(j)}$ is given by
 \begin{align}
     \label{eq:omegacorot}
        \Delta \omega_\pm^{(j)}
        &= \Delta\omega_\pm (t_0) + \Delta \alpha_\pm \, (t_j-t_0) ,
  \end{align}
where the superscript (C) is omitted henceforth since all calculations are performed in the co-moving frame, and the initial frequency $\Delta\omega_\pm (t_0)$ is related to the lab-frame value $\Delta\omega_\pm^\mathrm{(L)}(t_0)$ as follows:
  \begin{equation}
  \label{eq:omegaknotdefinition}
      \Delta\omega_\pm (t_0) = \Delta\omega_\pm^\mathrm{(L)} (t_0) \mp \mathbf{k}_\mathrm{eff}\cdot \mathbf{v}_0 .
  \end{equation}
with $\mathbf{v}_0 = \mathbf{v} (t_0)$. In turn, the laser phase $\varphi_\pm^{(j)}$ corresponds to
  \begin{equation}
    \label{eq:phicorot}
        \varphi_\pm^{(j)} = \varphi_{\pm} (t_0) + \Delta \omega_{\pm} (t_0)\,  (t_j-t_0) + \frac{1}{2} \Delta \alpha_{\pm} \, (t_j-t_0)^2 ,
  \end{equation}
where the initial phase in the co-moving frame coincides with the lab-frame value: $\varphi_{\pm} (t_0) = \varphi_{\pm}^\mathrm{(L)}(t_0)$.\\
\\
Note that the term proportional to $\mathbf{v} (t_0)$ in Eq.~\eqref{eq:omegaknotdefinition}, and its associated contribution to Eq.~\eqref{eq:phicorot}, leads to a dependence on the initial velocity of the phase shift $\delta \phi$ for unbalanced interferometers. This can happen for unequal times between the mirror pulse and the two beam-splitter pulses through the contributions of the laser phases $ \varphi_\pm^{(j)}$. It can also arise through the terms proportional to $\delta_\pm^{(\mathrm{c})} (t_j) /\Omega_\mathrm{D}^{(j)}$ when the intensities for both beam-splitter pulses (and hence $\Omega_\mathrm{D}^{(j)}$) are not exactly the same.
\subsection{Effect between pulses and dependence on \(t_j\)}
\label{subsec:chirpbetweenpulses}
The effects of $\Delta \alpha_\pm$ between pulses are captured by Eqs.~\eqref{eq:omegacorot} and \eqref{eq:phicorot}.
They explicitly depend on the times $t_j$, which are connected to the particular choice of mnemonic trajectories.
As emphasized in Sec.~\ref{sec:tj}, observable quantities (including the total phase shift $\delta \phi$) should be independent of the actual choice made for the times $t_j$. This was shown there to be the case in the absence of frequency chirps and accelerations, but for $\Delta \alpha_\pm \neq 0$ the argument needs to be generalized. In fact, the effect that any uncompensated frequency chirp between pulses has on $\Delta \omega_\pm^{(j)}$ and $\varphi_\pm^{(j)}$ involves terms proportional to $\Delta \alpha_\pm$ with an explicit dependence on $t_j$.
It turns out that this dependence is canceled out by an analogous dependence that arises when considering the effects of $\Delta \alpha_\pm$ during the pulses.
A perturbative calculation of such effects will  be provided in the next subsection, but the key aspects relevant for the present discussion can be explained here.\\
\\
The calculation in Sec.~\ref{subsec:chirpduringpulse} involves several time integrals, and 
for pulse envelopes that are symmetric with respect to the central time $\bar{t}_j$, it is convenient to introduce the time variable $t - \bar{t}_j$. Thus, before changing from the Schr\"odinger picture to the co-rotating frame, one first rewrites the term quadratic in time that appears in Eq.~\eqref{eq:phase-chirp} as
\begin{align}
\frac{1}{2} \Delta \alpha_\pm\, (t - t_j)^2 &= \frac{1}{2} \Big[ \Delta \alpha_\pm\, (t - \bar{t}_j)^2 + 2\, (\Delta \alpha_\pm \Delta t_j) (t - t_j)
\nonumber \\
&\quad\quad\ - \Delta \alpha_\pm \Delta t_j^2 \Big]
\label{eq:alpha_quadratic2} ,
\end{align}
with $\Delta t_j = \bar{t}_j - t_j$. The first term on the right-hand side, quadratic in $(t - \bar{t}_j)$, will be 
treated perturbatively in the next section. The term linear in $(t - t_j)$, on the other hand, can be combined with the analogous term in Eq.~\eqref{eq:phase-chirp} and amounts to the following redefinition of $\Delta \omega_\pm^{(j)}$:
\begin{equation}
\Delta \omega_\pm^{(j)} \to \Delta \omega_\pm^{(j)} + (\Delta \alpha_\pm \Delta t_j) = \Delta\omega_\pm (\bar{t}_j)
\equiv \Delta \bar{\omega}_\pm^{(j)}
\label{eq:omega-bar} ,
\end{equation}
where he have introduced a bar to denote quantities evaluated at time $\bar{t}_j$.\\
\\
One can then proceed as done in Sec.~\ref{sec:single_pulse} for $\Delta \alpha_\pm = 0$ and obtain analogous results but with the replacement $\delta_j \to \bar{\delta}_j$ that follows from the redefinition in Eq.~\eqref{eq:omega-bar}. Some of the phase terms proportional to the detuning are also proportional to time and depend on $t_j$. Such a dependence cancels out in the total phase shift $\delta\phi$, but the argument provided in Sec.~\ref{sec:fullAI} needs to be slightly adapted because of the time dependence of $\Delta\omega_\pm$. Indeed, if one writes in this case $t_j$ as $t_j = \bar{t}_j - \Delta t_j$, the term $\Delta \bar{\omega}_\pm^{(j)} \Delta t_j$ that arises from the dependence on $\Delta \bar{\omega}_\pm^{(j)}$ of the detuning $\bar{\delta}_j$ needs to be combined with the last term on the right-hand side of Eq.~\eqref{eq:alpha_quadratic2}. The result can then be absorbed in the following redefinition of the phase $\varphi_\pm^{(j)}$:
\begin{align}
\varphi_\pm^{(j)} &\to \varphi_\pm^{(j)} + \Delta \bar{\omega}_\pm^{(j)} \Delta t_j - \frac{1}{2} \Delta \alpha_\pm \Delta t_j^2 
\nonumber \\
&\quad\ = \varphi_\pm^{(j)} + \Delta \omega_\pm^{(j)} \Delta t_j + \frac{1}{2} \Delta \alpha_\pm \Delta t_j^2 \nonumber \\
&\quad\ = \varphi_\pm (\bar{t}_j) \equiv \bar{\varphi}_\pm^{(j)}
\label{} .
\end{align}
In this way, no explicit dependence on $\Delta t_j$ is left, and the total phase shift is manifestly independent of the particular choice for the times $t_j$. In fact, the result can be directly obtained by evaluating all the quantities at the times $t_j = \bar{t}_j$. Nevertheless, it was important to have freedom 
in the choice of times $t_j$. For example, by choosing  $t_j = t_j^*$ for the two beam-splitter pulses, the mnemonic trajectories coincide, at the end of each pulse, with the actual central trajectories of the atomic wave packets. This choice is particularly convenient when analyzing under what circumstances one has an open interferometer as well as the associated sensitivity to initial conditions.
\subsection{Effect during each pulse}
\label{subsec:chirpduringpulse}
The effect of $\Delta\alpha_\pm$ during each pulse will be calculated perturbatively. As detailed in Appendix~\ref{app:chirp}, the Hamiltonian is transformed to the co-rotating frame via the unitary operator $\hat{U}_j(t)$ (Eq.~\eqref{eq:unitaryrotation}), expressed in terms of $\Delta\bar{\omega}_\pm^{(j)}$ (Eq.~\eqref{eq:omegacorot}) and $\bar{\varphi}_\pm^{(j)}$ (Eq.~\eqref{eq:phicorot}), yielding the Hamiltonian in the co-rotating frame
  \begin{equation}
  \label{eq:Hamiltonianrotatingchirp}
      \hat{H}^\mathrm{(rot)} \approx \hat{H}_0^\mathrm{(rot)} + \frac{1}{2} \Omega_\mathrm{D} \Delta\alpha_\mathrm{A} (t - \bar{t}_j)^2  \hat{J}_2 \,,
  \end{equation}
where, coming from Eq.~\eqref{eq:chirpHintlong}, we consider only the antisymmetric component of the uncompensated chirp, $\Delta\alpha_\mathrm{A}$, as the symmetric part cancels out in a symmetric interferometer setup, and $\hat{H}_0^\mathrm{(rot)}$ is given by
  \begin{equation}
  \label{eq:Hamiltonian0chirp}
      \hat{H}_0^\mathrm{(rot)} = (\bar{\delta}_\mathrm{A} + \delta') \hat{J}_3 + \Omega_\mathrm{D} \hat{J}_1 \,.
  \end{equation}
  Changing to the interaction picture through the relation $\hat{H}_\mathrm{I} = \hat{U}^\dagger_{0}(t,\bar{t}_j) \hat{H}_\mathrm{int}^\mathrm{(rot)}(t) \hat{U}_{0} (t,\bar{t}_j)$ requires the unperturbed propagator
  \begin{align}
  \label{eq:propagator}
      &\hat{U}_0(t, \bar{t}_j) = e^{-\frac{i}{\hbar} \hat{H}_0^\mathrm{(rot)} (t - \bar{t}_j)} \notag \\
      &= \mathbf{I} - \frac{i}{\hbar} \hat{J}_1 \sin\left(\Omega_\mathrm{D} (t-\bar{t}_j)\right) + \frac{1}{\hbar^2} \hat{J}_1^2 (\cos\left(\Omega_\mathrm{D} (t-\bar{t}_j)\right) - 1) \,.
  \end{align}
  This closed form applies specifically to the spin-1 representation (Sec.~\ref{sec:intromethods}, Appendix~\ref{app:SU2}). Evaluating the Dyson series at first order yields
  \begin{equation}
      \hat{U}_\mathrm{I} \approx 1 - i \frac{2}{\hbar} \hat{J}_2 \sin\left(\mathcal{A}/2\right) f(\tau) \,,
  \end{equation}
  where the time integral with integrand $\hat{H}_\mathrm{I} \propto \hat{U}_0^\dagger(t, \bar{t}_j) \hat{J}_2 \hat{U}_0(t, \bar{t}_j)$ is simplified by changing to the time variable $(t - \bar{t}_j)$, which is particularly convenient for pulse envelopes that are symmetric around the central time $\bar{t}_j$. \\
  \\
  The chirp-induced evolution is captured by the time-dependent factor
  \begin{equation}
    \label{eq:ftau}
      f(\tau) = \Delta\alpha_\mathrm{A} \tau^2 \left(\frac{1}{8} - \frac{1}{\mathcal{A}^2} + \frac{1}{2\mathcal{A}} \cot\frac{\mathcal{A}}{2}\right) \,.
  \end{equation}
  This expression is valid for both single and double diffraction (see Appendix~\ref{app:chirp}). The complete time evolution in the Schr\"odinger picture is given by:
  \begin{align}
      |\psi (t) \rangle_\mathrm{S} &= e^{-\frac{i}{\hbar} E_0 \cdot \mathbb{I} \, (t-t_\mathrm{i})} \hat{U}^\dagger_j(t) \times \notag \\
      & \times \hat{U}_0(t, \bar{t}_j) \hat{U}_\mathrm{I} (t, t_\mathrm{i}) \hat{U}^\dagger_0(t_\mathrm{i}, \bar{t}_j) \hat{U}_j(t_\mathrm{i}) |\psi (t_\mathrm{i}) \rangle_\mathrm{S}\,,
  \end{align}
  with the following result for the product of unitary operators that arises when transforming back from the interaction picture:
  \begin{align}
  \label{eq:productofpropagators}
      \hat{U}_0(t, \bar{t}_j) \hat{U}_\mathrm{I} (t, t_\mathrm{i}) &\hat{U}^\dagger_0(t_\mathrm{i}, \bar{t}_j) \approx \mathbf{I} - \frac{i}{\hbar} \left(\hat{J}_1 + \hat{J}_2 f(\tau)\right) \sin\mathcal{A} \notag \\
       &   + \frac{1}{\hbar^2} \left(\hat{J}_1 + \hat{J}_2 f(\tau)\right)^2 (\cos\mathcal{A}-1) \,.
  \end{align}
  This derivation is performed perturbatively to first order in $f(\tau)$. In completing the square, a term proportional to $f(\tau)^2 \hat{J}_2^2$ arises, which is neglected as a higher-order contribution because both the pulse duration $\tau$ and the uncompensated chirp $\Delta\alpha_\mathrm{A}$ are small compared to the dominant contribution to the evolution. Physically, the finite-pulse chirp acts as a small rotation $\hat{J}_1 \to \hat{J}_1 + \hat{J}_2 f(\tau)$, which amounts to a small phase correction $\bar{\varphi}_\pm^{(j)}\to \bar{\varphi}_\pm^{(j)} \pm f(\tau)$ for the diffracted component, as shown in Eq.\,\eqref{eq:chirpinclusionofF}. For pulse envelopes or chirp profiles $\Delta\alpha(t)$ that are not parity even around $\bar{t}_j$, the first-order effect introduces an additional phase in the undiffracted state, as demonstrated for spin-\(1/2\) in Eq.\eqref{eq:asymmetricchirpspin12}.

\subsection{General result for a full atom interferometer}
  Combining the results from the preceding subsections, the total phase shift Eq.\,\eqref{eq:fullAIphasediff} extends to $\delta \phi = \delta \bar{\phi}_\mathrm{las} + \delta \bar{\phi}_\mathrm{det} + \delta \bar{\phi}_\mathrm{chirp}$. The subscript ``chirp'' denotes contributions arising from the finite pulse duration, whereas the laser and detuning terms capture between-pulse chirp effects. \\
  \\
  In the freely falling frame co-moving with the mid-point trajectory, the dominant laser phase contribution $\delta \bar{\phi}_\mathrm{las}$ (Eq.~\eqref{eq:phidet}) is evaluated at the reference time $\bar{t}_j$. The detuning $\bar{\delta}_\mathrm{A}^{(\mathrm{c})}$ is now generalized to incorporate the uncompensated chirp $\Delta \alpha_\mathrm{A}$, following the definition in Eq.~\eqref{eq:detuning} and the chirp-dependent frequency shift in Eq.~\eqref{eq:omegacorot}. \\
  \\
  Similarly, the phase contributions capturing the dynamics during the pulse, $\delta \bar{\phi}_\mathrm{det}$ (Eq.~\eqref{eq:phidet}), are evaluated at $\bar{t}_j$ and incorporate the uncompensated chirp $\Delta \alpha_\mathrm{A}$. This term accounts for any pulse asymmetry between the first and third beam-splitter pulses. Incorporating the finite-pulse-duration effects derived in Sec.~\ref{subsec:chirpduringpulse}, the uncompensated chirp during the pulses introduces an additional phase shift:
  \begin{equation}
      \delta \bar{\phi}_\mathrm{chirp} = - 2 \Delta \alpha_\mathrm{A} \left(\left(\tau_1^2 + \tau_3^2\right) c_1 - 2 \tau_2^2 c_2 \right) \,.
  \end{equation}
   For ideal $\pi/2$ and $\pi$ pulses, the coefficients evaluate to $c_1 = 1/8 - 4/\pi^2 + 1/\pi$ and $c_2 = 1/8 - 1/\pi^2$ respectively.\\
   \\
  Assembling all contributions, the total phase difference reads:
  \begin{widetext}
    \begin{align}
      \delta \phi &= - 2 \Delta\omega_\mathrm{A} (t_0) \, (\bar{t}_1 - 2 \bar{t}_2 + \bar{t}_3)  + \delta \varphi_0+ \delta_\mathrm{A}^\mathrm{(c)} (t_0) \, \tau_1 - \delta_{\mathrm{A}}^\mathrm{(c)} (t_0) \, \tau_3 -  2\, \delta_\mathrm{A}^{(\mathrm{c})}(t_0)\, \tau_1 \frac{ \tan (\mathcal{A}_1/2)}{\mathcal{A}_1} + 2\, \delta_\mathrm{A}^{(\mathrm{c})}(t_0)\, \tau_3 \frac{ \tan (\mathcal{A}_3/2)}{\mathcal{A}_3} \notag \\
      -& 2 \Delta \alpha_\mathrm{A} \left[ \frac{1}{2}(\bar{t}_1^2 - 2 \bar{t}_2^2   + \bar{t}_3^2) + \left( \bar{t}_1 \frac{\tau_1}{2} - \bar{t}_3 \frac{\tau_3}{2} \right) - \left( \bar{t}_1 \tau_1  \frac{ \tan (\mathcal{A}_1/2)}{\mathcal{A}_1} - \bar{t}_3 \tau_3  \frac{ \tan (\mathcal{A}_3/2)}{\mathcal{A}_3} \right)   + \Big(\!\left(\tau_1^2 + \tau_3^2\right) c_1 - 2 \tau_2^2 c_2 \Big) \right]
      \label{eq:total_phase_shift} ,
  \end{align}
  \end{widetext}
   where $\delta \varphi_0$ accounts for the initial laser phases according to Eq.~\eqref{eq:initiallaserphase} and the discussion thereafter, $\mathcal{A}_j = \Omega_\mathrm{D}^{(j)} \tau_j$ are the pulse areas, and we have taken the time origin at $t_0$. The general result for arbitrary time origin can be strightforwardly obtained by replacing \(\bar{t}_j \to \bar{t}_j - t_0\) throughout the equation. In addition, it is often convenient to introduce the times between pulses $T_1 = \bar{t}_2 - \bar{t}_1$ and $T_2 = \bar{t}_3 - \bar{t}_2$, and to take $t_0 = \bar{t}_1$. Note that while identical pulse durations eliminate the first two terms that depend on $\delta_{\mathrm{A}}^\mathrm{(c)} (t_0)$ in Eq.~\eqref{eq:total_phase_shift}, intensity fluctuations can still induce non-trivial phase shifts through the remaining detuning contributions. \\
    \\
  For a symmetric Mach-Zehnder configuration with $\tau_1 = \tau_3 = \tau_2/2 = \tau$, $\mathcal{A}_1 = \mathcal{A}_3 = \mathcal{A}_2/2 = \pi/2$, and interrogation times $T_1 = T_2 = T$, the phase shift simplifies to
  \begin{align}
      \delta \phi &= \delta \varphi_0 - 2\, \big( \alpha_\mathrm{A} - \mathbf{k}_\mathrm{eff}\cdot \mathbf{a} \big) \, \times \notag \\ 
      & \quad \times \left[T^2  + T \tau \left( \frac{4}{\pi}-1\right) + \tau^2 \left( \frac{2}{\pi} - \frac{3}{4}\right) \right] \,.
  \end{align}
This result generalizes to double diffraction the finite-pulse corrections derived in Ref.~\cite{Antoine_2006}. Note that our definition of the interrogation time $T$ measures the interval between pulse centers, whereas Ref.~\cite{Antoine_2006} defines it between pulse edges.

\section{Additional effects}
\label{sec:additional_effects}
This section deals with additional effects beyond the simpler case of constant accelerations and uniform gravitational fields explicitly considered in the previous section.

\subsection{Time-dependent acceleration and frequency chirp}
\label{sec:time-dependent_chirp}
In the previous sections we have given explicit results for the case of time-independent accelerations and frequency chirps. For time-dependent ones, on the other hand, the explicit form of $\alpha_\pm (t)$ and $\mathbf{a} (t)$ needs to be taken into account when calculating the time integrals in Eqs.~\eqref{eq:vel} and \eqref{eq:omegawithchirp1} as well as the additional time integral relating the frequency difference $\Delta\omega_\pm (t)$ to the laser phase $\varphi_\pm (t)$. In this respect, it is 
convenient to explicitly consider sinusoidal functions with angular frequency $\omega$, in terms of which general time-dependent functions can be decomposed through a Fourier transform. The uncompensated frequency chirp 
$\Delta\alpha_\pm (t)$ is then written as
\begin{equation}
\Delta\alpha_\pm (t) =\Delta\tilde{\alpha}_\pm^\mathrm{(c)} (\omega) \, \cos \omega ( t - \bar{t}_2 )
+ \Delta\tilde{\alpha}_\pm^\mathrm{(s)} (\omega) \, \sin \omega ( t - \bar{t}_2 )
\label{eq:t-dep_alpha} ,
\end{equation}
where we have chosen $\bar{t}_2$, the central time of the $\pi$ pulse, as time reference%
\footnote{This choice is convenient for the subsequent calculations, but the results can be straightforwardly generalized to an arbitrary time $t_0$, instead of $\bar{t}_2$, with the following replacements: $\Delta\tilde{\alpha}_\pm^\mathrm{(c)} (\omega) \to \Delta\tilde{\alpha}_\pm^\mathrm{(c)} (\omega) \, \cos \omega ( \bar{t}_2 - t_0 ) - \Delta\tilde{\alpha}_\pm^\mathrm{(s)} (\omega) \, \sin \omega ( \bar{t}_2 - t_0 )$ and $\Delta\tilde{\alpha}_\pm^\mathrm{(s)} (\omega) \to \Delta\tilde{\alpha}_\pm^\mathrm{(s)} (\omega) \, \cos \omega ( \bar{t}_2 - t_0 ) - \Delta\tilde{\alpha}_\pm^\mathrm{(c)} (\omega) \, \sin \omega ( \bar{t}_2 - t_0 )$. The choice $t_0 = \bar{t}_2$ also means that one should take $\Delta\omega_\pm (t_0) = \Delta\omega_\pm (\bar{t}_2)$ for the first term on the right-hand side of Eq.~\eqref{eq:total_phase_shift}, but the result can be generalized to an arbitrary $t_0$ with the substitution $\Delta\omega_\pm (\bar{t}_2) \to \Delta\omega_\pm (t_0) + \int_{t_0}^{\bar{t}_2} dt \, \Delta \alpha_\pm (t)$.
}. For any given angular frequency $\omega$, the obtained phase-shift result can then be compared to the approach based on the so-called sensitivity function \cite{Cheinet_2008}, which is often expressed in Fourier space.\\
\\
The leading contribution, of zeroth order in the pulse duration, can be directly obtained from the result for the laser phases that follows from 
performing a double time integral of Eq.~\eqref{eq:t-dep_alpha}.
The corresponding phase shift is thus given by
\begin{align}
\delta \tilde{\phi}_\mathrm{lead} (\omega) &= \big( 2 / \omega^2 \big)\, \Delta\tilde{\alpha}^\mathrm{(c)}_\mathrm{A} (\omega)\, \Big( \! \cos (\omega\, T_1) + \cos (\omega\, T_2) - 2 \Big)
\nonumber \\
&= - \big( 8 / \omega^2 \big)\, \Delta\tilde{\alpha}^\mathrm{(c)}_\mathrm{A} (\omega)\, \sin^2 \big( \omega\, T / 2 \big)
\label{} ,
\end{align}
where the second equality holds for the case $T_1 = T_2 = T$.
Note that the result for the time-independent case is recovered when taking the limit $\omega \to 0$
and identifying the constant parameter $\Delta\alpha_\mathrm{A}$ with $\Delta\tilde{\alpha}^\mathrm{(c)}_\mathrm{A} (0)$.\\
\\
From the new results for $\Delta\omega_\pm (t_j)$ and the corresponding detunings $\delta_\pm (t_j)$, one can obtain the phase-shift corrections linear in the pulse duration, which amount to the following replacement for the terms in Eq.~\eqref{eq:total_phase_shift} that are linear in the pulse duration and $\Delta\alpha_\mathrm{A}$:
\begin{equation}
\Delta\alpha_\mathrm{A}\, \big( \bar{t}_j - t_0 \big) 
\, \to \, \int_{t_0}^{\bar{t}_j} dt \,\, \Delta\alpha_\mathrm{A} (t)
\label{} ,
\end{equation}
where the expression is valid for a general $t_0$. In contrast, Eq.~\eqref{eq:t-dep_alpha} is valid when choosing the intermediate time of the $\pi$ pulse as time reference, which corresponds to taking $t_0 = \bar{t}_2$. Nevertheless, it can be straightforwardly generalized to an arbitrary $t_0$ by proceeding as described in the corresponding footnote.\\
\\
Finally, the time-dependent chirp rate in Eq.~\eqref{eq:t-dep_alpha} will also give rise to $\omega$-dependent
corrections to the effects of 
frequency chirp and acceleration during the pulses, which correspond to the terms in Eq.~\eqref{eq:total_phase_shift} that are quadratic in $\tau_j$.
The detailed calculation is analogous to the derivation of Appendix~\ref{app:chirp} for the time-independent case. For each pulse, labeled by the index~$j$, one should first rewrite the right-hand side of Eq.~\eqref{eq:t-dep_alpha} as a liner combination of sine and cosine functions with argument $\omega ( t - \bar{t}_j )$.
Similarly to the computation for constant $\Delta\alpha_\mathrm{A}$, the contribution of the cosine term, which is an even function of $( t - \bar{t}_j )$, leads to a small phase correction for the diffracted wave packet. In contrast, the sine term, which is an odd function, gives rise (also) to a small correction for the phase of the undiffracted wave packet.
Corrections to the amplitude, on the other hand, arise at higher orders in perturbation theory \cite{Jiang_2026}.

\subsection{Gravity gradients}
Generalizing our results beyond uniform fields to include the effects of gravity gradients, which are characterized by the tensor $\Gamma_{ij} = \partial^2 U / \partial x^i \partial x^j$, is relatively straightforward. Indeed, for gravity gradients that are approximately uniform over length scales comparable to the maximum arm separation, the mid-point trajectory still corresponds to a freely falling trajectory. In the comoving frame, the central trajectories of the two interferometer arms are no longer straight lines due to tidal effects, but they are still symmetric around the mid-point. Therefore, both the propagation phases  as well as the spatially dependent part of the laser phases are, in this frame, the same for both arms and result in a vanishing net contribution to the phase shift.\\
\\
The non-trivial phase contributions associated with gravity gradients come from the spatially independent laser phase terms that arise when transforming from the lab frame to the freely falling one. They can be obtained by considering the expressions for the mid-point trajectory (and its time derivative) up to first order in $\Gamma_{ij}$:
\begin{align}
\bar{\mathbf{X}} (t) &= 
\big( \mathbf{X}_0 + \mathbf{v}_0 \, \Delta t \big) + \frac{1}{2} \mathbf{a}\, \Delta t^2
\nonumber \\
&\ \ + \frac{1}{2} \big( \Gamma \, \Delta t^2 \big) \left( \mathbf{X}_0
+ \frac{1}{3} \mathbf{v}_0 \, \Delta t + \frac{1}{12} \mathbf{a}\, \Delta t^2 \right)
\label{eq:gg2} ,
\end{align}
and
\begin{align}
\mathbf{v}(t) = d \bar{\mathbf{X}} / dt &= \mathbf{v}_0 + \mathbf{a}\, \Delta t
\nonumber \\
&\ \ + \Gamma \left( \mathbf{X}_0 + \frac{1}{2} \mathbf{v}_0 \, \Delta t
+ \frac{1}{6} \mathbf{a}\, \Delta t^2 \right)
\label{eq:vel_gg} ,
\end{align}
where $\mathbf{X}_0 = \bar{\mathbf{X}} (t_0)$, $\mathbf{v}_0 = \mathbf{v} (t_0)$ and we have introduced $\Delta t = (t - t_0)$.
The result for $\Delta\omega_\pm$ at leading order in $\Gamma$ corresponds then to
\begin{align}
\Delta\omega_\pm (t) &= \Delta\omega_\pm^\mathrm{(L)} (t)
\mp \mathbf{k}_\mathrm{eff}\cdot \mathbf{v}(t)
\label{eq:doppler_gg} ,
\end{align}
with $\mathbf{v}(t)$ given by Eq.~\eqref{eq:vel_gg}.
Furthermore, by integrating again over time and evaluating the result at time $\bar{t}_j$, one gets the laser phase $\bar{\varphi}_j$ for each pulse. Adding up the contributions of the three pulses gives the following result for the leading phase-shift correction due to a time-independent gravity gradient:
\begin{align}
\delta \phi_\mathrm{gg} &=
\mathbf{k}_\mathrm{eff}^\mathrm{T}\, \big( \Gamma \, T^2 \big) \left( \mathbf{X}_0 + \mathbf{v}_0 \, T
+ \frac{1}{6} \mathbf{a}\, T^2 \right)
\label{eq:phase-shift_gg} ,
\end{align}
where $T$ is the time between pulses (for a balanced interferometer with $\bar{t}_2 - \bar{t}_1= \bar{t}_3 - \bar{t}_2 = T$), and we have taken as initial time $t_0$ the central time of the first pulse ($t_0 = \bar{t}_1$). [These results can be generalized to time-dependent gravity gradients by integrating once or twice over time the expression $\big( d^2 / dt^2 \big) \bar{\mathbf{X}}^{(1)} = \Gamma \Big( \mathbf{X}_0 + \mathbf{v}_0 \, \Delta t + (1/2)\, \mathbf{a}\, (\Delta t)^2 \Big)$ to obtain the first-order gravity-gradient correction to $\mathbf{v} (t)$ and $\bar{\mathbf{X}} (t)$, respectively.]\\
\\
At first order in the pulse durations, there are phase-shift contributions linear in $\Gamma$ and the pulse durations $\tau_j$ that arise from the antisymmetric detunings $\delta_\mathrm{A} (\bar{t}_j)$.
Indeed, Eqs.~\eqref{eq:vel_gg}--\eqref{eq:doppler_gg} imply gravity-gradient corrections to $\Delta\omega_\pm (\bar{t}_j)$ suppressed by factors of order $(\Gamma T^2)$,
and the corresponding changes of $\delta_\mathrm{A} (\bar{t}_j)$ contribute to the phase shift $\delta \phi$ through the terms proportional to $\delta_\mathrm{A} (\bar{t}_j)$ and $\tau_j$ in Eq.~\eqref{eq:total_phase_shift}.
In principle, there is also an effect due to the modified trajectories that result from the tidal effects associated with gravity gradients. However, in the comoving frame the modified trajectories for the two arms are symmetric around the mid-point, and the phases associated with the resulting corrections to the central detuning are the same for both arms.
(Contrary to the effects considered above for the transformation to the comoving frame, in this case the change of $\delta_\mathrm{A}$ has opposite signs for the two arms due to the opposite velocities induced by the tidal effects.)\\
\\
Higher-order corrections of order $(\Gamma \tau^2)$ correspond to the effects of the gravity gradients during the pulse, but also to differences in the chirp rate $\Delta \alpha_j$ for each pulse that result from the effects of gravity gradients between pulses.
The chirp rates $\Delta \alpha_\pm^{(j)} = d\, \Delta \omega_\pm / dt \, \big|_{\bar{t}_j}$ are obtained by taking the time derivative of Eqs.~\eqref{eq:vel_gg}--\eqref{eq:doppler_gg} and evaluating the result at the corresponding time $\bar{t}_j$ for each pulse:
\begin{align}
\Delta \alpha_\pm^{(j)} &= \alpha_\pm^\mathrm{(L)} \mp \mathbf{k}_\mathrm{eff}\cdot \bigg[ \, \mathbf{a}
+ \Gamma \, \Big( \, \mathbf{X}_0 + \mathbf{v}_0 \, \big( \bar{t}_j - t_0 \big) 
\nonumber \\
& \qquad \qquad \qquad \qquad \qquad \ + \frac{1}{2} \mathbf{a}\, \big( \bar{t}_j - t_0 \big)^2 \Big) \bigg]
\label{eq:alpha_j} ,
\end{align}
where the time dependence arises from the correction terms proportional to $\Gamma$, which also involve certain dependence on the initial position and velocity. Such gravity-gradient corrections to $\Delta \alpha_\pm^{(j)}$ contribute to the phase shift $\delta\phi$ through the terms proportional to $\Delta \alpha_\pm^{(j)} \tau_j^2$ on the in Eq.~\eqref{eq:total_phase_shift}.\\
\\
A detailed analysis of the effects of the gravity gradients during the pulse, lies beyond the scope of this paper and will be presented elsewhere. Nevertheless, some general aspects can be briefly outlined here. Firstly, from the symmetry of the two arm trajectories in the freely falling frame comoving with the mid-point, it is clear that the leading phase corrections will be the same for both arms, with no net contribution to $\delta\phi$. And similarly for any wave-packet distortions caused by the tidal effects during the pulse. Asymmetries between the two arms can, on the other hand, arise when simultaneously considering the effects of gravity gradients during the pulse and of the antisymmetric detuning  $\delta_\mathrm{A} (\bar{t}_j)$ that results from the change to the comoving frame, but they will be suppressed by higher powers of the pulse duration.\\
\\
Secondly, the effects of gravity gradients during the pulses give rise to a small correction of order $(\Gamma \, \tau^2)$ to the relative displacement (in position and momentum) between the interfering wave packets at each exit port. The correction is suppressed by a factor $(\tau / T)^2$ compared to the main contribution of order $(\Gamma \, T^2)$. Such open interferometers, which can be understood as a result of the tidal effects associated with gravity gradients \cite{Roura_2017}, exhibit a sensitivity to the initial conditions that is directly related to the relative displacements at the exit ports \cite{Roura_2014,Roura_2017}. (In fact, the small corrections to the displacement are connected with the phase-shift corrections discussed right after Eq.~\eqref{eq:alpha_j} that depend on the initial position and velocity.)
This sensitivity to initial conditions is a major source of systematic effects in high-precision measurements based on atom interferometry, but a very effective mitigation technique has been proposed \cite{Roura_2017} and successfully implemented \cite{D_Amico_2017,Overstreet_2018}. It involves a common shift of the laser frequencies for the mirror pulse (implying a small change of $\mathbf{k}_\mathrm{eff}$) by an amount that is determined through self-calibration. For tests of UFF, the corrections to this shift due to finite pulse duration
are implicitly accounted for in the self-calibration process \cite{Asenbaum2020}. On the other hand, in gradiometric measurements that require an accurate determination of the gravity gradient (such as a measurement of the gravitational constant $G$ \cite{Rosi_2018}), those corrections need to be properly taken into account.

\subsection{Asymmetric frequency chirp and retardation effects}
As shown in Ref.~\cite{Roura_2025}, relativistic effects including time dilation and retardation associated with the propagation of the laser pulses can be conveniently described in the freely falling frame comoving with the mid-point trajectory. In such a frame, these effects simply lead to shifts of light rays and laser wave fronts while keeping their slope unchanged (for propagation parallel or antiparallel to the velocity with respect to the lab frame).
The focus in that reference was on laser pulses driving single-photon transitions, but the approach can be extended to two-photon transitions and then the results obtained in Ref.~\cite{Tan_2017} are recovered. In the latter case, one needs to consider also the counter-propagating frequency component. For simplicity one can take a constant amplitude for one of the two frequency components ($\omega_2$)
and pulse the amplitude of the other component ($\omega_1$), 
which then determines the pulse envelope and timing.\\
\\
In two-photon processes, first-order Doppler shifts have opposite signs for the two counter-propagating frequency components (with opposite unit vectors $\hat{\mathbf{n}}_1 = - \hat{\mathbf{n}}_2$). Thus, the most natural way of compensating the effects of accelerations is an antisymmetric frequency chirp for both components. But alternative ways of achieving an equivalent compensation for the two-photon frequency difference $\Delta \omega$ are often employed. For example, applying a chirp with twice the rate to just one of the two frequency components. However, deviations from an antisymmetric frequency chirp give rise to non-trivial retardation effects on the laser phase that for single diffraction constitute the leading
phase-shift correction associated with retardation \cite{Gillot_2016,Tan_2017}.\\ 
\\
Experiments relying on double diffraction in atomic fountains \cite{Malossi_2010,Zhou_2015}, typically employ three laser frequency components: one common to both counter-propagating pairs and two whose frequencies can be chirped independently to guarantee that both counter-propagating pairs stay on resonance. The asymmetry in the frequency chirp for each pair leads to the leading retardation effects mentioned in the previous paragraph. Nevertheless, since these phase contributions remain unchanged under reversal of $\mathbf{k}_\mathrm{eff}$, in double diffraction they are the same for both arms, which have opposite momentum transfers for each pulse, and cancel out.
The asymmetric frequency chirps also lead to a change of the effective momentum transfer for each laser pulse. But the change is equal for both counter-propagating pairs ($\Delta \mathbf{k}_\mathrm{eff}^+ = \Delta \mathbf{k}_\mathrm{eff}^-$), and the effects cancel out for double diffraction (where $\mathbf{k}_\mathrm{eff}^+ = - \mathbf{k}_\mathrm{eff}^-$). Indeed, the total momentum transfer for the $\pi$ pulse corresponds to the linear combination $\mathbf{k}_\mathrm{eff}^- - \mathbf{k}_\mathrm{eff}^+$, which remains unchanged. And so does the momentum difference between the two arms generated by the beam-splitter pulses.
[Analogous conclusions apply to the much smaller effect due to the first-order Doppler shifts of the frequency components and implying $\Delta \mathbf{k}_\mathrm{eff}^+ = \Delta \mathbf{k}_\mathrm{eff}^- = - (\Delta \omega_\pm / c)\, ( \mathbf{v} / c)$.]\\
\\
Finally, in single diffraction retardation effects on the timing of the pulse envelopes give rise to additional phase-shift corrections \cite{Tan_2017}, which can be suppressed by adjusting the pule emission times so that it compensates the retardation effects. For double diffraction, this could be an option in the atom fountain set-ups involving two independent frequency components that contribute each to a different counter-propagating pair. On the other hand, it is not possible for experiments in microgravity (or horizontal configurations) with a retro-reflecting mirror. A more detailed analysis of the associated phase-shift corrections will be provided elsewhere, but one expects that most of them are invariant under a reversal of the momentum transfer direction and cancel out when taking the difference between the two arms in double diffraction. The only remaining contribution would be proportional to the detuning and the time shift of the pulse envelope for the $\pi$ pulse.\\
\\
Note that time dilation effects do not play a role in double Raman interferometry. In single diffraction, these effects are mainly connected to differences between the two arms of the proper times spent by the atoms in the two internal states \cite{Roura_2025}, and the corresponding phase terms are proportional to the energy difference $\Delta E$. However, compared to atom interferometers based on single-photon transitions, the effects are suppressed by about five orders of magnitude for Raman transitions, which involve transition frequencies $\Delta E / \hbar$ in the microwave rather than the optical regime.
And more importantly, in double diffraction there is no change of internal state along each arm and the state is the same on both arms.\\
\\
Thus, relativistic effects are altogether much less significant in double than in single diffraction, and their impact is limited to retardation effects on the timing of pulse envelopes proportional to the frequency detuning (of the $\pi$ pulse).
\subsection{Magnetic fields and other external forces}
An analysis of the effects of magnetic fields (and other external forces) within an approach where the frame comoving with the mid-point trajectory plays a central role can be found in Appendix~H of Ref.~\cite{Roura_2025}. As shown there, one of the main phase-shift contributions corresponds to the difference between the external potentials for the internal states of the two arms evaluated along the mid-point trajectory. This can be an important contribution for atom interferometers based on single Raman diffraction, even for homogeneous but time-dependent magnetic fields, because the internal state of the atoms at any given time is different for the two arms.\\
\\
In contrast, for double Raman diffraction the atoms in both arms are in the same internal state and this contribution vanishes identically. In fact, this feature constitutes an important advantage of double over single diffraction for interferometers based on Raman transitions.\\
\\
In case of spatial inhomogeneities, a common external potential for the atoms in both arms can still give rise to non-trivial effects due to modifications of the central trajectories of the atomic wave packets. For gradients that are uniform over length scales comparable to the arm separation, the effects can be entirely captured through the modification of the mid-point trajectory that results from the force associated with the gradient of the external potential. This modification implies that the frame comoving with the mid-point is no longer a freely falling frame. Nevertheless, it can be taken into account by adding the acceleration $- (\boldsymbol{\nabla} V_\mathrm{ext}) / m$ caused by the gradient of the external potential to the acceleration $\mathbf{a}$ appearing in all the above results.
And even non-uniform gradients of the potential can be accounted for by considering the results for gravity gradients and adding $(\partial^2  V_\mathrm{ext} / \partial x^i  \partial x^j ) / m$ to the gravity-gradient tensor $\Gamma_{ij}$.\\
\\
For weak magnetic fields and atoms in a magnetically sensitive state (with $m_F \neq 0$), the magnetic potential is dominated by the linear Zeeman term and given by $V_\mathrm{ext} (t, \mathbf{x}) = m_F g_\mathrm{L} \mu_\mathrm{B} \big| \mathbf{B} (t, \mathbf{x}) \big|$, where $\mu_\mathrm{B}$ is the Bohr magneton, $g_\mathrm{L}$ is the Land\'e factor for that state, and the quantization axis is defined by the direction of the magnetic field $\mathbf{B} (t, \mathbf{x})$, which is mainly determined by the applied bias field. On the other hand, for atoms in a magnetically insensitive state ($m_F = 0$) the magnetic potential is dominated by the quadratic Zeeman effect, with $V_\mathrm{ext} (t, \mathbf{x}) \propto \big| \mathbf{B} (t, \mathbf{x}) \big|^2$.\\
\\
Finally, note that although illustrated for magnetic fields, the discussion in this subsection also applies to other external forces such as the force due to temperature gradients and associated with the optical potential that arises through the AC Stark shift induced by black body radiation.
\subsection{AC Stark shifts and Gaussian beams}
\label{sec:AC-Stark-shift}
Each one of the frequency components in the Raman pulses induce an AC Stark shift of the energy of the two internal states involved in the Raman transition. Since the excited state is the same for both branches in double diffraction, the corresponding contributions to the detunings $\delta_+$ and $\delta_-$ are the same (implying $\delta_\mathrm{A} = 0$ and a non-vanishing $\delta_\mathrm{S}$).
For the first and final beam-splitter pulses, the associated phases are therefore equal for both arms, resulting in a vanishing net contribution to the phase shift. And for the mirror pulse, only the symmetric part of the detuning $\delta_\mathrm{S}$ leads to a non-vanishing phase correction in each arm. Hence, symmetry between the two arms would imply again a cancelation of the phase-shift contributions.\\
\\
The AC Stark shift for the two internal states depends on the laser intensity of the various frequency components. For Gaussian beams with finite width, these intensities can be slightly different for large spatial separations of the two interferometer arms along the beam direction.
A detailed assessment of how significant this effect can be depends on a number of parameters characterizing the specific configuration (such as the width of the beam waist, its distance to the retro-reflection mirror, and the positions of the atom clouds in the two arms when the $\pi$ pulse is applied), and lies beyond the scope of this work.\\
\\
In fact, throughout the paper we have implicitly considered plane waves for the laser pulses, which can be a good approximation for sufficiently small atom clouds compared to the beam size. 
The effects of Gaussian beams with finite width will be considered in a separate publication. Nevertheless, certain aspects can be already inferred from the results obtained here. For example, thermal clouds can be described in terms of an incoherent ensemble of pure Gaussian states corresponding to wave packets with a much smaller size. Thus, under suitalbe conditions each wave packet can be described with the treatment presented here, but with a different Rabi frequency (or, equivalently, pulse area) for each member of the ensemble. The fact that, due to thermal expansion, each one of these wave packets explores different parts of the beam profile at different times, is then straightforwardly taken into account by using the appropriate Rabi frequency for each pulse, which leads in general to open interferometers and can naturally account for the effects reported in Ref.~\cite{Morel_2020a}.

\subsection{Parasitic paths}
\label{sec:parasitic_paths}

Contrary to Bragg diffraction, a single pair of counter-propagating laser frequencies do not drive Raman transitions to other diffraction orders besides the resonant one. That is because they are suppressed by much larger detunings of several GHz, which correspond to the energy splitting between the two hyperfine states.
Compared to single diffraction, in double Bragg diffraction, there are additional off-resonant transitions that play a significant role. Most of them are absent in the Raman case for the same reason given above for single diffraction, but there is a small set of higher-order off-resonant transitions that do contribute. They correspond to four-photon processes with a momentum transfer of $\pm 2\, \hbar \mathbf{k}_\mathrm{eff}$ and no change of internal state%
\footnote{There are also eight-photon processes and even higher-order ones, but their amplitudes are very suppressed.}.
A detailed treatment that incorporates such processes involves a description in terms of a 7-level system \cite{Hartmann2020,Hartmann2020a}, which has not been pursued here. Nevertheless, one expects that the corrections of those off-resonant processes to the phase accumulated along both arms will be the same because of the manifest symmetry between the two arms in the comoving frame.
On the other hand, such off-resonant processes also lead to parasitic paths (discussed in the next paragraph) that can give non-trivial contributions to the measured outcome at the exit ports.\\
\\
Indeed, off-resonant transitions give rise to diffracted wave-packet components that follow other central trajectories different from the two main arms and give additional contributions to the observed interference signal. Interferometers based on Raman diffraction involve far fewer parasitic paths compared to those employing Bragg pulses. Moreover, in double Raman diffraction the internal state for most spurious paths is different from the main two arms, and they can be removed with state-selective blow-away pulses after the first and second pulses \cite{Leveque_2009}.
Combined with spatially resolved detection of the exit ports, almost all parasitic paths can be eliminated. The only remaining ones are a pair of parasitic arms with central momenta $\pm 3\, \hbar \mathbf{k}_\mathrm{eff}$ and the same internal state as the main arms.
In this case, each pulse involves a six-photon process that is far detuned and results in an amplitude of the diffracted wave-packets suppressed by a factor of order $( \Omega^2 / 16\, \omega_\mathrm{rec}^2 )$.
And the final amplitude of the wave-packet components that follow those parasitic paths is suppressed by the cube of that factor. Therefore, its contribution to the interference signal will typically be negligible for the relatively long pulses with $\Omega \ll \omega_\mathrm{rec}$ that can be employed in experiments with ultracold atoms.

\section{Discussion}

In this work we have provided a comprehensive analytical treatment of the effects of pulse durations in double Raman interferometry, both in terms of atomic wave-packet evolution and of the interferometric phase shift.
Our study includes time-dependent accelerations and frequency chirps, gravity gradients, 
and retardation effects. AC Stark shifts induced by the Raman pulses and external forces such as those due to magnetic fields or temperature gradients in the black-body radiation have also been considered.
Altogether the methods developed and results obtained constitute a very valuable tool for high-precision measurements based on such kinds of atom interferometers in both ground- and space-based experiments.

In fact, accurate measurements in standard ground-based applications of atom interferometers to gravimetry (and inertial sensing) rely on scanning the chirp rate for the frequency difference between the two counter-propagating laser beams, which greatly relaxes the need for an accurate knowledge of the interferometer times and the effects of finite pulse durations.
In contrast, such an approach is not possible for experiments in microgravity employing double diffraction with a retro-reflection set-up. Therefore, a detailed characterization of the systematic effects associated with finite pulse durations, as provided here,
is crucial for achieving the required accuracy in absolute acceleration measurements such as those targeted by the CARIOQA mission \cite{Leveque2023}.

Our analytical calculations are valid for atomic wave packets with a sufficiently small momentum spread.
(Note that very short pulses should not be used so as to avoid losses to off-resonant diffraction orders and contributions from the parasitic paths discussed in Sec.~\ref{sec:parasitic_paths}.)
We have considered a momentum width \(\sigma_\mathrm{p} = 0.02\, \hbar \mathbf{k}_\mathrm{eff}\) and checked the validity of our approximation with numerical simulations of the wave-packet evolution during the pulses (with a pulse duration of $100~\mu\text{s}$). Such narrow momentum distributions are anyway required in experiments involving long interferometer times of several seconds and can be achieved through the combined use of ultracold atoms (possibly BECs) and matter-wave lensing techniques \cite{Muentinga_2013,Kovachy2015,Kovachy_2015,Xiong_2026}. 
Thus, any phase-shift corrections due to the finite momentum width of the atomic wave packets will be rather small. Especially, because their net contribution to the total phase shift is substantially suppressed due to the symmetry between the two interferometer arms in the frame comoving with the mid-point trajectory.
A detailed quantitative analysis of this point will be provided in future work.

There are two additional aspects that we also plan to investigate further in future work.
The first one is the impact of finite beam size in realistic Gaussian beams as well as the change of the laser intensity along the beam direction. The latter point can be particularly relevant for the mirror pulse because differences in the laser intensities for the two spatially separated arms can lead to different AC Stark shifts, which result in unequal symmetric detunings $\delta_\mathrm{S}$ for the two arms. And as discussed in Sec.~\ref{sec:AC-Stark-shift}, this would imply that the corresponding phase-shift contributions no longer cancel out.
The second aspect has to do with polarization imperfections of the laser pulses (concerning the polarization of the two injected frequency components or the quarter-wave plate in front of the retro-reflection mirror).
Such imperfections can give rise to additional optical lattices
that drive spurious Bragg transitions. Any resulting parasitic paths can be eliminated through the combination of blow-away pulses and spatially resolved detection, as explained in Sec.~\ref{sec:parasitic_paths}. But the additional optical lattices can also lead to a significant decrease of the diffraction efficiency, especially for the mirror pulse \cite{Jenewein2023}, and optimization strategies should be pursued.
Moreover, in the final beam-splitter the spurious Bragg transitions lead to systematic effects in the phase shifts at the two side ports. In principle, these effects could be effectively suppressed when adding up the phase shifts separately obtained from the outcome of each side port, as considered in Sec.~\ref{sec:fullAI}, but one still expects a contrast reduction for the overall interferometric signal, and the whole issue deserves a more detailed study.

\section*{Acknowledgments}
The authors thank Nico Schwersenz for a careful reading of the manuscript and for his valuable comments.
N.A.\ is also grateful to Matthias Meister for his support in layout and planning.

\newpage
\appendix

\section{Atom-light interaction}
\subsection{Three-level system}
\label{app:ALthreelevel}
The Hamiltonian in the Schr\"odinger picture, after adiabatic elimination of the excited state (see \cite{Steck2020, Bernhardt_1981, Marte_1992,Brion_2007}), under the dipole and rotating wave approximations, is given by
\begin{equation}
\label{eq:hamiltonianschrodi}
  \hat{H} = E_0 \mathbf{1} + \hat{H}_{\Delta E} + \hat{H}_\mathrm{int} \,,
\end{equation}
where
\begin{equation}
\label{eq:hamiltonianshrodideltaE}
  \hat{H}_{\Delta E} =  \Delta E_\mathrm{S} \frac{\hat{J}_3^2}{\hbar^2}+ \Delta  E_\mathrm{A} \frac{\hat{J}_3}{\hbar} \,,
\end{equation}
is expressed in terms of the symmetric and antisymmetric frequency differences: $\Delta  E_\mathrm{S}  = (E_+ + E_-)/2 - E_0$ and $\Delta  E_\mathrm{A}  = (E_+ - E_-)/2$, along with the generator $\hat{J}_3$ defined in Appendix~\ref{app:SU2}.  The basis states are represented as column vectors: 
\begin{align}
  | 1 \rangle  \to \begin{pmatrix} 0 \\ 1 \\ 0\end{pmatrix}, \quad | 2^+ \rangle \to \begin{pmatrix} 1 \\ 0 \\ 0 \end{pmatrix}  , \quad  | 2^- \rangle \to  \begin{pmatrix} 0 \\ 0 \\ 1 \end{pmatrix}  \,.
\end{align}
The interaction Hamiltonian reads
\begin{equation}
H_\mathrm{int}= \hbar \Omega(t)
\begin{pmatrix}
0 &  e^{-i \phi_+(t)}  & 0 \\
 e^{i \phi_+(t)}   & 0 & e^{i \phi_-(t)}  \\
0 &  e^{-i \phi_-(t)}  & 0
\end{pmatrix} \,,
\end{equation}
with the phases
\begin{align}
  \label{eq:phaseingredientsphirs}
      \phi_+(t) &= \varphi^{(j)}_+ + \Delta \omega^{(j)}_+ (t - t_j) \notag \\
      \phi_-(t) &= \varphi^{(j)}_- + \Delta \omega^{(j)}_- (t - t_j) \,,
\end{align}
depending on the laser phase and frequency at time $t_j$ defined in Eq.\,\eqref{eq:phij} and\,\eqref{eq:omegaj}.\\
 \\
The unitary operator
  \begin{equation}
  \label{eq:unitaryrotation}
      U_j(t) = \begin{pmatrix} e^{i \phi_+(t) } & 0 & 0 \\
        0  & 1 & 0 \\
        0 &  0 & e^{i \phi_-(t)} 
    \end{pmatrix} \,,
  \end{equation}
  transforms to the co-rotating frame, yielding the Hamiltonian in Eq.~\eqref{eq:rotatingHamiltonian}. This transformation relies on the detunings expressed in terms of their symmetric and antisymmetric components, defined in Eqs.~\eqref{eq:detuningsymm} and~\eqref{eq:detuningantisymm}, which are rewritten as $\delta_\mathrm{S} = \Delta E_\mathrm{S}/\hbar - \Delta\omega_\mathrm{S}$ and $\delta_\mathrm{A} = \Delta E_\mathrm{A}/\hbar - \Delta\omega_\mathrm{A}$, respectively. In a symmetric interferometer, the symmetric detuning $\delta_\mathrm{S}$ cancels out in the interferometric phase, so we set $\delta_\mathrm{S} \to 0$; it may be included perturbatively if needed. Note that the term proportional to $\delta_\mathrm{S}$ involves $\hat{J}^2$, which is not a linear combination of the SU(2) generators, its form depends on the representation (trivial for spin-1/2, where $\hat{J}^2 \propto \mathbb{I}$, but non-trivial and outside the SU(2) algebra for spin-1).\\
  \\
  The time evolution operator is rotated back via $\hat{U}_j(t)$, so that the time evolution in the Schr\"odinger picture reads
  \begin{equation}
      \hat{U}(t,t_\mathrm{i}) = e^{-\frac{i }{\hbar} E_0 \, (t-t_\mathrm{i})}  \hat{U}_j^\dagger(t) \, \hat{U}^\mathrm{(rot)}(t,t_\mathrm{i}) \, \hat{U}_j(t_\mathrm{i}) \,, 
  \end{equation}
  to describe the time evolution from initial state $s$ to final state $r$:
  \begin{equation}
       \langle  r |\Psi(t)\rangle 
    = \sum_s   \,  \Bigl<  r\Big|\hat{U}(t,t_\mathrm{i}) \Big| s \Bigr> \langle  s|\Psi(t_\mathrm{i})\rangle  \,,
  \end{equation}
  where phase terms arise from the rotation back, depending on the internal state:
  \begin{equation}
      \hat{U}_j(t) | s \rangle = e^{i \phi_{s}(t)} | s \rangle \,, 
  \end{equation}
  which relates the matrix elements of the evolution operators in the two frames via an additional phase factor:
  \begin{equation}
      \Big< r \Big| \hat{U}(t,t_\mathrm{i}) \Big|s \Big> 
    = \phi_{rs}^\mathrm{(full)} \, \Big<  r \Big|   \hat{U}^\mathrm{(rot)}(t,t_\mathrm{i})  \Big| s \Big> \,, 
  \end{equation}
  with the total phase
  \begin{equation}
  \label{eq:generalphirs}
      \phi_{rs}^\mathrm{(full)} = - \frac{E_0}{\hbar} (t-t_\mathrm{i}) - \phi_{r} (t_\mathrm{f}) +  \phi_{s} (t_\mathrm{i}) \,,
  \end{equation}
  with the phase contributions $\phi_{r}$ and $\phi_{s}$ defined in Eqs.\,\eqref{eq:phaseingredientsphirs} and $\phi_0 = 0$.\\
  \\
  The phases can be rewritten by expressing the laser frequency in terms of energy and residual detuning
   \begin{equation}
   \label{eq:frequencydetuning}
      \Delta \omega^{(j)}_\pm = \frac{E_\pm}{\hbar} - \frac{E_0}{\hbar} -  \delta_\pm \,,
  \end{equation}
  leading to the organization in terms of mnemonic trajectories. Substituting Eqs.\,\eqref{eq:phij} and\,\eqref{eq:omegaj} into Eq.\,\eqref{eq:phaseingredientsphirs} confirms that the resulting phases are independent of $t_j$.
\subsection{Including center-of-mass motion}
\label{app:COM}
The atom-light interaction entangles internal states with the center-of-mass motion. The state is described as a tensor product of continuous momentum eigenstates and normalized internal states. We consider a three-level system, neglecting higher-order or off-resonant couplings. The results from the previous subsection are applied with the following identifications:
  \begin{align}
      | 1 \rangle  &\to | \mathbf{p} , \mathrm{g} \rangle  \notag \\
      | 2_+ \rangle &\to| \mathbf{p} + \hbar \mathbf{k}_\mathrm{eff}, \mathrm{e} \rangle \notag \\
      | 2_- \rangle & \to | \mathbf{p} - \hbar \mathbf{k}_\mathrm{eff}, \mathrm{e} \rangle  \,.
  \end{align}
  In the main text, these are abbreviated as  $|0\rangle$, $|+\rangle$, and $|-\rangle$, respectively.
  For this mapping to hold, the relative shape of the wave packet during coupling must be preserved, meaning the deviation $\mathbf{p}'$ from the central momentum remains unchanged:
  \begin{equation} 
    \label{eq:deltainmomentum}
        \langle \mathbf{p}'_\mathrm{f} |\mathbf{p}'_\mathrm{i} \rangle = 
    \delta^{(3)} \big( \mathbf{p}'_\mathrm{f}-\mathbf{p}'_\mathrm{i} \big) = \frac{1}{(2 \pi \hbar)^3} \int d^3x \, e^{-\frac{i}{\hbar} (\mathbf{p}'_\mathrm{f} -\mathbf{p}'_\mathrm{i})\cdot \mathbf{x}} \,.
\end{equation}
The energies including the center-of-mass motion are
  \begin{align}
  \label{eq:energiesCOM}
      E_0 &= E_\mathrm{g} +  \frac{\mathbf{p}^2}{2m}  \notag \\
      E_+ &= E_\mathrm{g}  + \hbar \omega_\mathrm{eg} +  \frac{(\mathbf{p} + \hbar \mathbf{k}_\mathrm{eff})^2}{2m}  \notag \\
      E_- &= E_\mathrm{g} + \hbar \omega_\mathrm{eg} +  \frac{(\mathbf{p} - \hbar \mathbf{k}_\mathrm{eff})^2}{2m} \,,
  \end{align}
  with the transition frequency 
  \begin{equation}
       \omega_\mathrm{eg} = (E_\mathrm{e} - E_\mathrm{g})/\hbar \,,
  \end{equation}
   the recoil frequency 
  \begin{equation}
      \omega_\mathrm{rec} = \frac{\hbar \mathbf{k}_\mathrm{eff}^2}{2m} \,,
  \end{equation}
    and the Doppler frequency
  \begin{equation}
      \omega_\mathrm{D} = \frac{\mathbf{p} \cdot \mathbf{k}_\mathrm{eff}}{m} \,.
  \end{equation}
  The symmetric and antisymmetric energy differences are then expressed as $\Delta  E_\mathrm{S} = \hbar (\omega_{\mathrm{eg}}+ \omega_{\mathrm{rec}})$ and $\Delta  E_\mathrm{A} = \hbar \omega_{\mathrm{D}}$. The momentum is split into a central component $\mathbf{P}$ and deviation $\mathbf{p}'$ from the center as $\mathbf{p} = \mathbf{P} + \mathbf{p}'$ (Eq.\,\eqref{eq:momentumdecomposition}), and the detuning is decomposed accordingly as given in Eq.\,\eqref{eq:detuningdecomposed}. The laser frequency decomposition in Eq.\,\eqref{eq:frequencydetuning} is refined to depend only on the central energy and central detuning:
  \begin{equation}
  \label{eq:centraldet}
      \delta_\pm^{(\mathrm{c})}  = \frac{E_\pm^{(\mathrm{c})}}{\hbar} - \frac{E_0^{(\mathrm{c})}}{\hbar} -  \Delta \omega^{(j)}_\pm \,,
  \end{equation}
  The central detuning vanishes when the laser frequency is resonant with the center of the wave packet. The central energy $E_0^{(\mathrm{c})}$ is the $\mathbf{p}'$-independent part of Eq.~\eqref{eq:energiesCOM}. To obtain the central energies for all three states, evaluate Eq.~\eqref{eq:energiesCOM} by replacing $\mathbf{p}$ with $\mathbf{P}_\mathrm{g}$. This leads to the phase accumulation in Eq.\,\eqref{eq:generalphirs}, explicitly for all transitions: 
    \begin{align}
    \phi_\mathrm{gg}&=  -\frac{E_0^{(\mathrm{c})}}{\hbar} (t_\mathrm{f}-t_\mathrm{i})\notag \\
    \phi_\mathrm{\pm \pm} &=  \left( -\frac{E_\pm^{(\mathrm{c})}}{\hbar} + \delta^{(\mathrm{c})}_\pm  \right)(t_\mathrm{f}-t_\mathrm{i}) \notag \\
    \phi_\mathrm{g\pm} &= -\frac{E_0^{(\mathrm{c})}}{\hbar} (t_\mathrm{f}-t_j) - \left( \frac{E_\pm^{(\mathrm{c})}}{\hbar} -\delta_\pm^{(\mathrm{c})}  \right)(t_j-t_\mathrm{i})+ \varphi^{(j)}_\pm \notag \\
    \phi_\mathrm{\pm g}&=   -\frac{E_0^{(\mathrm{c})}}{\hbar} (t_j - t_\mathrm{i}) - \left(  \frac{E_\pm^{(\mathrm{c})}}{\hbar} - \delta^{(\mathrm{c})}_\pm \right)(t_\mathrm{f}-t_j)-   \varphi^{(j)}_\pm \notag \\
    \phi_\mathrm{\pm \mp}&=  - \left(\frac{E_\pm^{(\mathrm{c})}}{\hbar} -\delta_\pm^{(\mathrm{c})} \right) (t_\mathrm{f}-t_j) \notag \\
    &- \left( \frac{E_\mp^{(\mathrm{c})}}{\hbar} +\delta_\mp^{(\mathrm{c})} \right)(t_j-t_\mathrm{i}) \mp  \varphi^{(j)}_+ \pm \varphi^{(j)}_- \,.
    \end{align}
    Here, the organization of the phase terms according to mnemonic trajectories becomes explicit: the phase accumulates according to the initial energy up to the reference time $t_j$, and thereafter according to the final energy. For the upper arm of a Mach-Zehnder interferometer, the total phase is obtained by summing the individual contributions: $\phi_{+g}$ from the first beam splitter, $\phi_{-+}$ from the mirror, and $\phi_{g-}$ from the final beam splitter.\\
    \\
    Adding and subtracting phase terms proportional to the initial and final kinetic energy $\pm \mathbf{P}_s^2/(2m\hbar)(t_j - t_\mathrm{i})$ and $\pm \mathbf{P}_r^2/(2m\hbar)(t_\mathrm{f} - t_j)$ yields the decomposition $\phi_{rs} \to \mathcal{S}_{rs} + \phi^{(j)}_{rs}$. Here, $\mathcal{S}_{rs}$ is the classical action from Eq.~\eqref{eq:classicalaction}, and $\phi^{(j)}_{rs}$ is the corrective phase factor listed in the left column of Table~\ref{tab:phaseterms}. To facilitate the transformation of the plane-wave factor, we define the position $\mathbf{X}_s^{(j)}$ at the reference time $t_j$ as \(\mathbf{X}_s^{(j)} = \mathbf{X}_\mathrm{i} + \mathbf{P}_s/m \cdot(t_j - t_\mathrm{i})\)  for the trajectory in state $s$. A detailed discussion of the phase reorganization and the role of the definition of \(\mathbf{X}_s^{(j)}\) is provided in Ref.~\cite{Augst2027}.
\subsection{Evolution in position space}
\label{app:Grot}
The time evolution operator from the time-independent Hamiltonian in the co-rotating frame (Eq.\,\eqref{eq:rotatingHamiltonian}) yields the propagator
\begin{equation}
\label{eq:Grotdefinition}
    \Big< \mathbf{p}_\mathrm{f}, r \Big| \hat{U}^\mathrm{(rot)}(t,t_\mathrm{i}) \Big|\mathbf{p}_\mathrm{i}, s \Big> 
=  G_{rs}^\mathrm{(rot)}\big(\mathbf{p}', t, t_\mathrm{i} \big) \,
\delta^{(3)} \big( \mathbf{p'_\mathrm{f}}-\mathbf{p'_\mathrm{i}} \big) \,,
\end{equation}
which is defined this way because the matrix elements of the unitary evolution operator are proportional to a Dirac delta function, as given in Eq.\,\eqref{eq:deltainmomentum}. Using the generalized Euler formula (Eq.\,\eqref{eq:eulerspin1}) with the Hamiltonian in the co-rotating frame Eq.\eqref{eq:rotatingHamiltonian}, the time evelution for a square pulse is determined by the pulse area $\mathcal{A} = \Omega_\mathrm{eff} (t - t_\mathrm{i})$. In this case the effective Rabi frequency (Eq.~ \eqref{eq:omegaeff}) is expressed solely in terms of the antisymmetric detuning $\delta_\mathrm{A}$, and the rotation vector is given by
    \begin{equation}
        \Omega_\mathrm{eff} = \sqrt{\Omega_\mathrm{D}^2 + \delta^2_\mathrm{A}}, \quad
        \mathbf{n} = \frac{1}{ \Omega_\mathrm{eff}(t)}\begin{pmatrix}
    \Omega_\mathrm{D} \\ 
    0\\ 
    \delta_\mathrm{A} \end{pmatrix} \,,
    \end{equation}
    where we introduce the generalized Rabi frequency of the three-level system \(\Omega_\mathrm{D} = \sqrt{2}\Omega\) and obtain the unitary evolution operator $\hat{U}^\mathrm{(rot)}(t,t_\mathrm{i})$. Using Eq.~\eqref{eq:Grotdefinition}, we then derive the propagator
    \begin{widetext}
        \begin{equation}
            G_{rs}^\mathrm{(rot)}\big( \mathbf{p}', t, t_\mathrm{i} \big)  = \frac{1}{ \Omega_\mathrm{eff}^2} 
            \begin{pmatrix}
            \Omega^2  & -\delta_\mathrm{A} \Omega  & -\Omega^2  \\
            -\delta_\mathrm{A} \Omega   & \delta_\mathrm{A}^2 & \delta_\mathrm{A} \Omega \\
            -\Omega^2  & \delta_\mathrm{A} \Omega   & \Omega^2 
            \end{pmatrix}
             -  \frac{i}{ \Omega_\mathrm{eff}} \begin{pmatrix}
            \delta_\mathrm{A} & \Omega  & 0 \\
            \Omega  & 0 & \Omega  \\
            0 & \Omega    & -\delta_\mathrm{A}
            \end{pmatrix}
             \sin\mathcal{A}
            + \frac{1}{ \Omega_\mathrm{eff}^2} \begin{pmatrix}
            \delta_\mathrm{A}^2 + \Omega^2  & \delta_\mathrm{A} \Omega  & \Omega^2  \\
            \delta_\mathrm{A} \Omega   & 2 \Omega^2 & -\delta_\mathrm{A} \Omega \\
            \Omega^2  & -\delta_\mathrm{A} \Omega   & \delta_\mathrm{A}^2 + \Omega^2 
            \end{pmatrix}
             \cos\mathcal{A} \,,
        \end{equation}
    \end{widetext}
    which agrees with the expression derived in Ref.\,\cite{Giese2013} (up to a sign convention, where the double Bragg diffraction in the deep Bragg regime is treated as a three-level system in the reference, comparable to double Raman diffraction). Note that \(G_{rs}^\mathrm{(rot)} \big( \mathbf{p}', t, t_\mathrm{i} \big) \) also depends on the central momentum \(\mathbf{P}_\mathrm{i}\) through its dependence on the detuning (Eq.\,\eqref{eq:centraldet}). This dependence is implicit and is therefore not included explicitly among the arguments.\\
    \\
  Rearranging the terms and applying double-angle trigonometric identities, we neglect higher-order detuning contributions ($\delta_\mathrm{A}^2 \to 0$). 
  These corrections are small and mainly affect the amplitudes of the undiffracted components, leading to wave packet distortions (see Ref.\cite{Augst2027}).  Using the approximation $(1 + i \epsilon) \approx e^{i \epsilon}$ for $\epsilon \ll 1$ and the approximation of the effective Rabi frequency to$\Omega_\mathrm{D}$ (Eq.~\eqref{eq:approxomegaD}), we obtain the approximate propagator, following the approach of Ref.\cite{Augst2027}:
  \begin{widetext}
  \begin{align}
  \label{eq:propagatorapprox}
     & G_{rs}^\mathrm{(rot)}\big( \mathbf{p}', t, t_\mathrm{i} \big)  \approx  \frac{ \Omega }{\Omega_\mathrm{D}}
        \begin{pmatrix}
         \frac{  \Omega }{\Omega_\mathrm{D}} \left( 1 + \cos\mathcal{A}  \right) e^{-i  \frac{\Omega_\mathrm{D} \delta_\mathrm{A}}{\Omega^2} \tan\frac{\mathcal{A}}{2}}
         & 
        - i \sin\mathcal{A} \, e^{ -  i \frac{\delta_\mathrm{A}}{ \Omega_\mathrm{D}  } \tan \frac{\mathcal{A}}{2}}
        & 
         - \frac{ \Omega }{\Omega_\mathrm{D}} \left( 1 - \cos\mathcal{A}  \right)
        \\
         - i  \sin\mathcal{A} \, e^{-i  \frac{\delta_\mathrm{A}}{ \Omega_\mathrm{D} }\tan\frac{\mathcal{A}}{2}} 
        & 
        \frac{  2 \Omega   }{ \Omega_\mathrm{D}} \cos \mathcal{A}
        & 
         -  i \, \sin\mathcal{A} \, e^{i  \frac{\delta_\mathrm{A}}{ \Omega_\mathrm{D} }\tan\frac{\mathcal{A}}{2}} 
        \\
         \frac{  \Omega }{\Omega_\mathrm{D}}\left( 1 - \cos\mathcal{A}  \right)
        & 
        - i \sin \mathcal{A} \, e^{  i\frac{\delta_\mathrm{A}}{ \Omega_\mathrm{D}  } \tan \frac{\mathcal{A}}{2}} 
        & 
        \frac{ \Omega }{\Omega_\mathrm{D}} \left( 1 + \cos\mathcal{A}  \right)  e^{i  \frac{\Omega_\mathrm{D} \delta_\mathrm{A}}{\Omega^2} \tan\frac{\mathcal{A}}{2}}
        \end{pmatrix} \,. \\
        &\text{Following Ref.~\cite{Augst2027}, the wave packet evolution in position space reads} \notag \\
        &
        \langle \mathbf{x}, r |\Psi(t)\rangle 
        = \frac{1}{(2 \pi \hbar)^{3/2}} \,  \sum_s \, e^{ \phi_{rs}} e^{\frac{i}{\hbar} \mathbf{P}_\mathrm{f} \cdot \mathbf{x}} \, e^{- \frac{i}{\hbar} \mathbf{P}_\mathrm{i} \mathbf{X}_\mathrm{i}}  \int d^3p' \,e^{-\frac{i}{\hbar}\frac{\mathbf{p'}^2}{2m} (t-t_\mathrm{i})} \, e^{\frac{i}{\hbar} \mathbf{p'}  \cdot   \left( \mathbf{x} -\mathbf{X}_\mathrm{i} - \frac{\mathbf{P}_\mathrm{g}}{m} \cdot  (t-t_\mathrm{i}) \right)}\, G_{rs}^\mathrm{(rot)} \big( \mathbf{p}', t, t_\mathrm{i} \big) \, \psi^{(s)}_\mathrm{c}(\mathbf{p}',t_\mathrm{i}) \,,
        \label{eq:waveevogeneral}
  \end{align}
  \end{widetext}
    where $\mathbf{P}_{\mathrm{i/f}} \in \{ \mathbf{P}_\mathrm{g}, \mathbf{P}_\mathrm{g} \pm \hbar \mathbf{k}_\mathrm{eff} \}$ denote the initial and final central momenta. The quadratic and last linearly $\mathbf{p}'$-dependent phase terms originate from the rotation back to the Schr\"odinger picture, as encoded in the full phase $\phi_{rs}^\mathrm{(full)}$ [Eq.~\eqref{eq:generalphirs}], with $\phi_{rs}$ representing its $\mathbf{p}'$-independent component. All remaining plane-wave factors arise from the Fourier transform.\\
    \\
 According to the Fourier shift theorem, linear $\mathbf{p}'$-dependent phase terms induce a shift in the central trajectory. We therefore analyze how the phase contributions from $G_{rs}^\mathrm{(rot)}$ modify the plane-wave factor $\mathbf{p}' \cdot \left(\mathbf{x} - \mathbf{X}_{rs}(t)\right)$ and consequently shift the central trajectory $\mathbf{X}_{rs}(t)$ for different diffraction channels:
  \begin{itemize}
        \item The undiffracted ground state and fully diffracted excited states (anti-diagonal elements of Eq.~\eqref{eq:propagatorapprox}, $rs \in \{\mathrm{gg}, -+, +-\}$) acquire no additional linear $\mathbf{p}'$-dependent phase, and thus experience no extra trajectory shift. Their central trajectory is therefore
        \begin{equation}
            \mathbf{X}_\mathrm{gg} = \mathbf{X}_\mathrm{\mp\pm} = \mathbf{X}_\mathrm{i} + \frac{\mathbf{P}_\mathrm{g}}{m} \cdot (t-t_\mathrm{i}) \,,
        \end{equation}
        as detailed in Eqs.~\eqref{eq:trajXgg} and~\eqref{eq:trajXdiffe}. In the main text, we employ a compact single-subscript notation, as the initial state $s$ is unambiguously determined by the context (ground state for the first beam splitter, Sec.~\ref{sec:firstBS}; upper/lower excited states for the mirror and second beam splitter, Sec.~\ref{sec:mirrorandsecondBS}).
      \item For transitions between the ground and excited states (the four off-diagonal cross-coupling elements of Eq.~\eqref{eq:propagatorapprox} with $rs \in \{\mathrm{g}+,+\mathrm{g}\}$ and $rs \in \{\mathrm{g}-,-\mathrm{g}\}$), the central trajectory is given by
        \begin{equation}
            \mathbf{X}_\mathrm{g\pm} = \mathbf{X}_\mathrm{\pm g} = \mathbf{X}_\mathrm{i} + \frac{\mathbf{P}_\mathrm{g}}{m} \cdot (t-t_\mathrm{i}) \pm \Delta \mathbf{X}(t-t_\mathrm{i}) \,,
        \end{equation}
        with the trajectory shift $\Delta \mathbf{X}$ defined in Eq.\,\eqref{eq:deltaX}. Diffraction to and from the upper and lower excited state shifts the trajectory by $+\Delta \mathbf{X}(t-t_\mathrm{i})$  and $-\Delta \mathbf{X}(t-t_\mathrm{i})$, respectively, as detailed in Eqs.~\eqref{eq:trajXgepm} and~\eqref{eq:trajXepmtog}.
      \item For the undiffracted excited states (the corner elements of the main diagonal in Eq.~\eqref{eq:propagatorapprox}, $rs \in \{++\}$ and $rs \in \{--\}$), applying the approximation from Eq.~\eqref{eq:approxomegaD} yields
        \begin{equation}
            \mathbf{X}_\mathrm{\pm\pm} \approx \mathbf{X}_\mathrm{i} + \frac{\mathbf{P}_\mathrm{g}}{m} \cdot (t-t_\mathrm{i}) \pm 2\Delta \mathbf{X}(t-t_\mathrm{i}) \,.
        \end{equation}
        This corresponds to a central trajectory shift of $+ 2\Delta \mathbf{X}$ for the upper and $-2\Delta \mathbf{X}$ for the lower excited state, as detailed in Eq.~\eqref{eq:trajfullyidffracted}.
  \end{itemize}
  The \(\mathbf{p}'\)-independent phase contributions of the propagator \(G_{rs}^\mathrm{(rot)}\) in Eq.~\eqref{eq:propagatorapprox}, denoted by \(\phi^\mathrm{(sq)}_{rs}\), are listed in the right column of Table~\ref{tab:phaseterms}. The \(\mathbf{p}'\)-dependent phase of \(G_{rs}^\mathrm{(rot)}\) shifts the central trajectory, as described above. Aligning the momentum-independent plane-wave factor with this shifted trajectory gives rise to additional spatially dependent phase corrections of the form \(\phi^\mathrm{(sq)}_{rs} \propto \mathbf{P}_r \cdot \Delta\mathbf{X}_{rs}(t_\mathrm{i}, t_j)/\hbar\), where
  \begin{align}
  \label{eq:omittedspatiallydepphaseterms}
      &\Delta\mathbf{X}_\mathrm{\pm \pm}(t_\mathrm{i}) &&= 2\Delta \mathbf{X}(t-t_\mathrm{i}) - \mathbf{v}_\mathrm{rec}(t-t_\mathrm{i})  \notag \\
    &\Delta\mathbf{X}_\mathrm{g \pm}(t_\mathrm{i}, t_j) &&= \Delta \mathbf{X}(t-t_\mathrm{i}) - \mathbf{v}_\mathrm{rec}(t_j-t_\mathrm{i})  \notag \\
    &\Delta\mathbf{X}_\mathrm{\pm g}(t_\mathrm{i}, t_j) &&= \Delta \mathbf{X}(t-t_\mathrm{i}) - \mathbf{v}_\mathrm{rec}(t-t_j)  \notag \\
    &\Delta\mathbf{X}_\mathrm{\pm \mp}(t_\mathrm{i}, t_j) &&= \mathbf{v}_\mathrm{rec}(t-t_j) - \mathbf{v}_\mathrm{rec}(t_j-t_\mathrm{i}) \,.
  \end{align}
  These shifts represent the difference between the mnemonic and actual trajectories, depending on the choice of $t_j$. For the main diffraction paths in an atom interferometer, these shifts vanish when $t_j$ is chosen at $t_j \in \{t_1^*, \bar{t}_2, t_3^*\}$, as defined in Eq.~\eqref{eq:tstar}. At these points, the mnemonic trajectory matches the actual path, leading to individual cancellation of the correction phase in each arm. In general, even when $t_j$ is not chosen at these optimal points, the phase contributions remain identical in both arms due to symmetry and thus cancel out globally in the differential interferometric phase. For this reason, these terms are not listed in Table~\ref{tab:phaseterms}, as they do not contribute to the total interferometer phase.

\section{Uncompensated frequency chirp during square pulse}
\label{app:chirp}
To compute the phase contribution from the uncompensated chirp during the pulse, we begin with the Hamiltonian in the Schr\"odinger picture (Eq.~\eqref{eq:hamiltonianschrodi}), and rewrite the energy difference term from Eq.~\eqref{eq:hamiltonianshrodideltaE} as
  \begin{equation}
      \hat{H}_{\Delta E}=\omega_{\mathrm{D}} \hat{J}_3 -  (\omega_{\mathrm{eg}}+ \omega_{\mathrm{rec}}) \hat{J}_3^2 / \hbar \,,
  \end{equation}
  and the interaction Hamiltonian including the uncompensated chirp
  \begin{widetext}
  \begin{equation}
      H_\mathrm{int} =  \hbar \Omega(t) 
        \begin{pmatrix}
        0 & e^{-i \left( \frac{1}{2} \Delta \alpha_+ \, \left(t - \bar{t}_j \right)^2 +  \bar{\varphi}^{(j)}_+ + \Delta \bar{\omega}^{(j)}_+  (t - \bar{t}_j) \right)}  & 0 \\
        e^{i \left( \frac{1}{2} \Delta \alpha_+ \, \left(t - \bar{t}_j \right)^2  +  \bar{\varphi}^{(j)}_+ + \Delta \bar{\omega}^{(j)}_+ (t - \bar{t}_j) \right)}   & 0 & e^{i \left( \frac{1}{2} \Delta \alpha_- \,\left(t - \bar{t}_j \right)^2 +  \bar{\varphi}^{(j)}_- + \Delta \bar{\omega}^{(j)}_- (t - \bar{t}_j) \right)}  \\
        0 &  e^{-i \left(\frac{1}{2} \Delta \alpha_- \, \left(t - \bar{t}_j \right)^2+  \bar{\varphi}^{(j)}_- + \Delta \bar{\omega}^{(j)}_- (t - \bar{t}_j)  \right)}  & 0
        \end{pmatrix} \,.
  \end{equation}
  \end{widetext}
Throughout this work, we use the following notation: Hamiltonians in the Schrödinger picture are denoted without a superscript, such as $\hat{H}_\mathrm{int}$, $\hat{H}_{\Delta E}$, or simply $\hat{H}$. In the co-rotating frame, the superscript $(\mathrm{rot})$ is used: $\hat{H}^\mathrm{(rot)}$, $\hat{H}_0^\mathrm{(rot)}$, $\hat{H}_\mathrm{int}^\mathrm{(rot)}$. In the interaction picture, the Hamiltonian is denoted by $\hat{H}_\mathrm{I}$.\\
\\
Transforming to the co-rotating frame via the unitary operator $\hat{U}_j(t)$ (Eq.~\eqref{eq:unitaryrotation}), now expressed in terms of $\Delta\bar{\omega}^{(j)}_\pm$ (Eq.\eqref{eq:omegacorot}) and $\bar{\varphi}^{(j)}_\pm$ (Eq.\eqref{eq:phicorot}), gives the Hamiltonian
  \begin{align}
      \hat{H}^\mathrm{(rot)} &=\hat{U}_j (t) \left(\hat{H}_{\Delta E} + \hat{H}_\mathrm{int} \right) \hat{U}_j^\dagger(t) + i \hbar (\partial_t \hat{U}_j(t)) \hat{U}_j^\dagger(t) \notag\\
      &=  \hat{H}_0^\mathrm{(rot)} + \hat{H}_\mathrm{int}^\mathrm{(rot)} \,,
  \end{align}
  where $\hat{H}_0^\mathrm{(rot)}$ is given in Eq.\,\eqref{eq:Hamiltonian0chirp}, and the interaction Hamiltonian in the co-rotating frame becomes
  \begin{widetext}
  \begin{align}
  \label{eq:chirpHintlong}
      H_\mathrm{int}^\mathrm{(rot)} &= \frac{i}{2} \Omega \hbar (t -\bar{t}_j)^2 \left[\begin{pmatrix}
        0 &   - \Delta \alpha_\mathrm{S}  & 0 \\
          \Delta \alpha_\mathrm{S}  & 0 &  \Delta \alpha_\mathrm{S}  \\
        0 &    -\Delta \alpha_\mathrm{S}   & 0
        \end{pmatrix}
        +
        \begin{pmatrix}
        0 &   - \Delta \alpha_\mathrm{A}  & 0 \\
          \Delta \alpha_\mathrm{A}  & 0 &  -\Delta \alpha_\mathrm{A}  \\
        0 &    \Delta \alpha_\mathrm{A}   & 0
        \end{pmatrix}
         \right] \,,
  \end{align}
  \end{widetext}
   where $\Delta \alpha_\pm = (\Delta \alpha_\mathrm{S} \pm \Delta \alpha_\mathrm{A})/2$. The symmetric contribution $\Delta \alpha_\mathrm{S}$ is not contributing in the overall interferometric phase and is neglected here (but could be included perturbatively). This leads to the Hamiltonian in Eq.\,\eqref{eq:Hamiltonianrotatingchirp}.\\
   \\
We transition to the interaction picture via $\hat{H}_\mathrm{I} = \hat{U}_{0}^\dagger (t,\bar{t}_j) \hat{H}_\mathrm{int}^\mathrm{(rot)} (t) \hat{U}_{0} (t,\bar{t}_j)$, employing the propagator $\hat{U}_{0} (t,\bar{t}_j)$ from Eq.\eqref{eq:propagator} for a time-independent Hamiltonian during a square pulse ($\Omega(t) = \Omega$). This expression is derived using the generalized Euler formula (Eq.\eqref{eq:eulerspin1}):
  \begin{align}
      \hat{H}_\mathrm{I} &=   \frac{1}{2} \Delta \alpha_\mathrm{A}  (t- \bar{t}_j)^2 \Omega_\mathrm{D} \, \hat{U}_{0}^\dagger  (t, \bar{t}_j)
           \hat{J}_2
        \hat{U}_{0}(t, \bar{t}_j) \,, 
  \end{align}
   For zero detuning ($\delta_\mathrm{A} + \delta' \to 0$), the evolution reduces to the rotation around the $\hat{J}_1$-axis by an angle $\Omega_\mathrm{D} (t-\bar{t}_j)$, rotating $\hat{J}_2$ into the $\hat{J}_2$–$\hat{J}_3$ plane:
  \begin{align}
  \label{eq:rotationUsJ2}
      &\hat{U}_0^\dagger(t, \bar{t}_j) \hat{J}_2 \hat{U}_0(t, \bar{t}_j)  \notag \\
      &= \hat{J}_2 \cos\left(\Omega_\mathrm{D} (t-\bar{t}_j)\right) - \hat{J}_3 \sin\left(\Omega_\mathrm{D} (t-\bar{t}_j)\right)  \,,
  \end{align}
  which can alternatively be obtained via direct matrix multiplication or the Baker–Campbell–Hausdorff formula.\\
  \\
 The propagator in the interaction picture, obtained to first order via the Dyson series, is given by $\hat{U}_\mathrm{I} \approx \mathbf{1} - \frac{i}{\hbar} \int_{t_\mathrm{i}}^t \hat{H}_\mathrm{I} (t') dt'$. The integral is then evaluated as
  \begin{widetext}
      \begin{equation}
          \int_{ -\frac{\tau}{2}}^{ \frac{\tau}{2}} \hat{H}_\mathrm{I} (u) du  =  \frac{1}{2 } \Delta \alpha_\mathrm{A}  \Omega_\mathrm{D}  \int_{ -\frac{\tau}{2}}^{ \frac{\tau}{2}} u^2  \, \left[\cos \left(\Omega_{\mathrm{D}} u \right) \,\hat{J}_2 - \sin\left(\Omega_{\mathrm{D}} u \right)\, \hat{J}_3 \right] du =  2 \hat{J}_2  \,\sin \left(\mathcal{A}/2\right) f(\tau) \,,
      \end{equation}
  \end{widetext}
  where $u = t' - \bar{t}_j$, $dt' = du$, and $\bar{t}_j = (t_\mathrm{f} - t_\mathrm{i})/2$ is chosen at the pulse center. The symmetry of the integrand allows cancellation of the odd term $u^2 \sin(\Omega_{\mathrm{D}} u)$, leaving only the even part. The factor $f(\tau)$ (defined in Eq.\,\eqref{eq:ftau}) is proportional to the uncompensated frequency chirp and quadratic in the pulse duration. The result of the integral coincides with the spin-1/2 case, where
    \begin{align}
    \label{eq:spin12integral}
       \int_{ -\frac{\tau}{2}}^{ \frac{\tau}{2}} \hat{H}_\mathrm{I} (u) du & =  \hbar \, \sigma_2 \sin \left(\mathcal{A}/2\right)f(\tau) \,.
  \end{align}
  From here, we compute the product in Eq.\,\eqref{eq:productofpropagators}:
  \begin{align}
      &\hat{U}_0(t, t_\mathrm{i}) \hat{U}_0(t_\mathrm{i}, \bar{t}_j) \, \left( 1  - i \frac{2}{\hbar} \hat{J}_2 \sin \left(\mathcal{A}/2\right) f(\tau) \right) \hat{U}^\dagger_0(t_\mathrm{i}, \bar{t}_j) \notag \\
      & =  \hat{U}_0(t, t_\mathrm{i}) - \frac{i}{\hbar} f(\tau) \left(\hat{J}_2 \sin\mathcal{A}+ \frac{i}{\hbar} \{\hat{J}_1, \hat{J}_2\} \left(\cos\mathcal{A} - 1\right)\right) \notag \\
      &\approx \mathbf{I} - \frac{i}{\hbar} \left(\hat{J}_1  + \hat{J}_2 f(\tau) \right) \sin\mathcal{A} \notag \\
      &\quad + \frac{1}{\hbar^2} \left(\hat{J}_1  + \hat{J}_2 f(\tau) \right)^2 (\cos\mathcal{A}-1) \,. 
  \end{align}
  The second line follows by applying the rotation of \(\hat{J}_2\), analogous to Eq.\,\eqref{eq:rotationUsJ2}, and expressing $\hat{U}_0(t, t_\mathrm{i})$ using Rodrigues' formula (Eq.\,\eqref{eq:rodriguesapp}). Performing the matrix multiplication applying the identities in Eqs.\,\eqref{eq:quadraticJ} and\,\eqref{eq:cubicJ} yields the anti-commutator $\{\hat{J}_1, \hat{J}_2\}$ in Eq.\,\eqref{eq:J1J2anticommutator}. This derivation is carried out perturbatively to first order in $f(\tau)$. The term $f(\tau)^2 \hat{J}_2^2$, added to complete the square in the third line, is of higher order and therefore negligible. For comparison, the spin-1/2 case is evaluated using the Pauli matrix properties, as outlined in Sec.~\ref{sec:intro}:
  \begin{align}
  \label{eq:rotationsigma2}
      &\hat{U}_0(t_\mathrm{f}, t_\mathrm{i}) \hat{U}_0(t_\mathrm{i}, t_j)  \sigma_2 \hat{U}^\dagger_0(t_\mathrm{i}, t_j) \notag\\ 
      &= \hat{U}_0(t_\mathrm{f}, t_\mathrm{i}) \left( \cos \frac{\mathcal{A}}{2} \, \sigma_2 -  \sin \frac{\mathcal{A}}{2}\, \sigma_3 \right) \sigma_2\sigma_2 \notag \\
    &= \hat{U}_0(t_\mathrm{f}, t_\mathrm{i}) \left(\cos \frac{\mathcal{A}}{2} \, + i \sin\frac{\mathcal{A}}{2} \,  \sigma_1 \right) \sigma_2 \notag \\
    &= \hat{U}_0(t_\mathrm{f}, t_\mathrm{i}) \hat{U}_0^\dagger(t_\mathrm{f}, t_\mathrm{i}) \, \sigma_2 = \sigma_2 \,,
  \end{align}
   leading to an additional term $\hat{U}_0(t, t_\mathrm{i}) \to \hat{U}_0(t, t_\mathrm{i}) - i f(\tau) \sigma_2$ due to the uncompensated chirp, in this case with no higher-order generator dependencies. \\
   \\
  As a final result, the uncompensated chirp appears in the phase $\phi_{s}$ discussed in Appendix\,\ref{app:COM} (Eq.\eqref{eq:phaseingredientsphirs}) as
  \begin{align}
  \label{eq:chirpinclusionofF}
      \phi_0 &= 0 \notag \\
      \phi_+(t) &= \bar{\varphi}^{(j)}_+ + \Delta \bar{\omega}^{(j)}_+(t - \bar{t}_j) + f(\tau) \notag \\
      \phi_-(t) &= \bar{\varphi}^{(j)}_- + \Delta \bar{\omega}^{(j)}_- (t - \bar{t}_j) - f(\tau) \,.
  \end{align}
 Therefore, in the first-order approximation, the symmetric component of the uncompensated chirp contributes an additional phase factor $f(\tau)$ to the laser phase: $\bar{\varphi}^{(j)}_\pm \to \bar{\varphi}^{(j)}_\pm \pm f(\tau)$. For the single-diffraction case with a general time-dependent chirp $\Delta\alpha(t)$ that is parity-odd (i.e., asymmetric), Eq.~\eqref{eq:spin12integral} extends to  
\begin{align}
    \int_{-\tau/2}^{\tau/2} &\hat{H}_\mathrm{I}(u) \, du \notag \\
    &= \hbar \left[ \sin\left(\frac{\mathcal{A}}{2}\right) \sigma_2 f_1(\tau) - \cos\left(\frac{\mathcal{A}}{2}\right) \sigma_3 f_2(\tau) \right] \,,
\end{align}
with $f_1(\tau)$ defined in Eq.~\eqref{eq:ftau} and $f_2(\tau)$ a function proportional to the chirp. The $\sigma_3$-dependent term can be rotated using the same approach as in Eq.~\eqref{eq:rotationsigma2} \( \hat{U}_0(t_\mathrm{f}, t_\mathrm{i}) \hat{U}_0(t_\mathrm{i}, t_j) \sigma_3 \hat{U}_0^\dagger(t_\mathrm{i}, t_j) = \sigma_3\). As a result, the leading-order correction to the undiffracted state is a phase shift of  
\begin{equation}
\label{eq:asymmetricchirpspin12}
    |\psi(t)\rangle \propto \cos\left(\frac{\mathcal{A}}{2}\right) e^{i f_2(\tau)} \mathbb{I} \, |\psi(t_\mathrm{i})\rangle.
\end{equation}
This result can be generalized to the spin-1 case in a similar manner.

\section{SU(2) representation: generators and useful identities}
\label{app:SU2}
The SU(2) generators in the fundamental representation (spin-1/2) are given by
  \begin{equation}
      \sigma_x = \begin{pmatrix}
        0 & 1 \\ 1 & 0
        \end{pmatrix}, \quad
        \sigma_y = \begin{pmatrix}
        0 & -i \\ i & 0
        \end{pmatrix}, \quad
        \sigma_z = \begin{pmatrix}
        1 & 0 \\ 0 & -1
        \end{pmatrix} \,.
  \end{equation}
The SU(2) generators in the adjoint representation (spin-1) are
  \begin{align}
      J_1 = \frac{\hbar}{\sqrt{2}} &\begin{pmatrix}
        0 & 1 & 0 \\
        1 & 0 & 1 \\
        0 & 1 & 0
        \end{pmatrix}, \quad
        J_2 = \frac{\hbar}{\sqrt{2}} \begin{pmatrix}
        0 & -i & 0 \\
        i & 0 & -i \\
        0 & i & 0
        \end{pmatrix}, \notag \\
        &\text{and}\quad
        J_3 = \hbar \begin{pmatrix}
        1 & 0 & 0 \\
        0 & 0 & 0 \\
        0 & 0 & -1
        \end{pmatrix} \,,
  \end{align}
  where the hat notation is omitted, as these expressions explicitly represent the basis-dependent matrices.\\
  \\
In the spin 1 case, combining the identity $(\hat{J}_3/\hbar)^3 = \hat{J}_3/\hbar$ and the rotation relation $n^l \hat{J}_l/\hbar = \hat{U}_0 (\hat{J}_3/\hbar) \hat{U}_0^\dagger$ with $\hat{U}_0 =  e^{-i \mathcal{A}\, n^l J_l/\hbar } $, the identity generalizes to arbitrary rotations: $(n^l \hat{J}_l/\hbar)^3 = n^l \hat{J}_l/\hbar$. This allows the extension of the generalized Euler formula:
  \begin{equation}
      e^{i\mathcal{A} \hat{J}_l/\hbar}  = I + i \hat{J}_l/\hbar \sin\mathcal{A} + \left(\hat{J}_l/\hbar\right)^2 (\cos\mathcal{A} -1) \,,
  \end{equation}
   to arbitrary rotations:
  \begin{equation}
  \label{eq:rodriguesapp}
      e^{-i \mathcal{A} \, n^l \hat{J}_l/\hbar }  = \mathbf{I}- i n^l \hat{J}_l/\hbar \sin\mathcal{A} + (n^l \hat{J}_l/\hbar)^2 (\cos\mathcal{A} -1) \,.
  \end{equation}
The symmetric detuning $\delta_\mathrm{S}$ as well as uncompensated chirp mismatch $\Delta \alpha_\mathrm{S}$ contribute to the Hamiltonian in the co-rotating frame proportional to $\hat{J}_3^2$, which is explicitly given by
  \begin{equation}
      \frac{J_3^2}{\hbar^2} = \begin{pmatrix}
        1 & 0 & 0 \\
        0 & 0 & 0 \\
        0 & 0 & 1
        \end{pmatrix} \,.
  \end{equation}
The anti-commutator relation $\{\hat{J}_1,\hat{J}_2\}$, used in App.\,\ref{app:chirp}, gives
  \begin{equation}
  \label{eq:J1J2anticommutator}
      \frac{\{J_1,J_2\}}{\hbar^2}
         = \begin{pmatrix}
        0 & 0 & -i \\
        0 & 0 & 0 \\
        i & 0 & 0 \,.
        \end{pmatrix} 
  \end{equation}  
In App.\,\ref{app:chirp}, we use the quadratic operator identities
\begin{align}
\label{eq:quadraticJ}
  \hat{J}_1 \hat{J}_2 &= \frac{1}{2}\{\hat{J}_1, \hat{J}_2\} + \frac{i\hbar}{2} \hat{J}_3, \notag \\
\hat{J}_1 \hat{J}_3 &= \frac{1}{2}\{\hat{J}_1, \hat{J}_3\} - \frac{i\hbar}{2} \hat{J}_2\,,
\end{align}
which are representation-independent and follow directly from decomposing any operator product into its symmetric and antisymmetric parts, $\hat{A}\hat{B} = \frac{1}{2}\{\hat{A}, \hat{B}\} + \frac{1}{2}[\hat{A}, \hat{B}]$, together with the angular-momentum commutation relations. In addition, we employ the spin-$1$ cubic operator identities
\begin{align}
\label{eq:cubicJ}
  \hat{J}_1^2 \hat{J}_2 &= \frac{\hbar^2}{2} \hat{J}_2 + \frac{i\hbar}{2} \{\hat{J}_1, \hat{J}_3\}, \notag \\
\hat{J}_1^2 \hat{J}_3 &= \frac{\hbar^2}{2} \hat{J}_3 - \frac{i\hbar}{2} \{\hat{J}_1, \hat{J}_2\} \,,
\end{align}
which follow from the characteristic spin-$1$ relation $(\hat{J}_1/\hbar)^3 = \hat{J}_1/\hbar$. Finally, the trigonometric identities used in App.\,\ref{app:chirp} are
  \begin{align}
      &\cos\mathcal{A} + 1 = 2 \cos^2\left(\frac{\mathcal{A}}{2}\right)  \notag \\
      &\cos\mathcal{A} - 1 = - 2 \sin^2\left(\frac{\mathcal{A}}{2}\right) \notag \\
      &\sin\mathcal{A} = 2 \sin\left(\frac{\mathcal{A}}{2}\right) \cos\left(\frac{\mathcal{A}}{2}\right) \,.
  \end{align}

\bibliography{apssamp}

\end{document}